\documentclass[preprint]{raa} 

\newcommand{\orcid}[1]{%
    \raisebox{0.7ex}{\scalebox{1}{
        \href{https://orcid.org/#1}{\includegraphics[height=1.5ex]{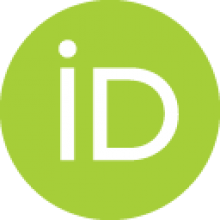}}%
    }}%
}

\usepackage{graphicx,times}             
\usepackage{natbib}
\usepackage{amssymb,amsmath}
\bibpunct{(}{)}{;}{a}{}{,}
\usepackage[pagebackref=true]{hyperref}

\begin{document}

  \title{\textit{SVOM}/VT: Instrument Overview, Science Objectives, and First-Year Performance
}

   \volnopage{Vol.0 (202x) No.0, 000--000}      
   \setcounter{page}{1}          

   \author{Yu-Lei Qiu \orcid{0009-0007-7207-4884}   \inst{1,*}\footnotetext{$*$Corresponding Authors, these authors contributed equally to this work.}
      \and Li-Ping Xin\orcid{0000-0002-9422-3437}
      \inst{1,*}   
         \and Jin-Song Deng\orcid{0000-0001-5646-8583}
        \inst{1,6,*}  
    \and Jian Zhang
    \inst{2}
    \and Xue-Wu Fan
    \inst{2}
        \and Hong-Bo Cai
        \inst{1}  
    \and Chao Wu\orcid{0009-0001-7024-3863}
      \inst{1}  
     \and Hua-Li Li
           \inst{1}
    \and Rui-Feng Su
    \inst{3}
    \and Qing-Yun Mao
        \inst{3}
    \and Wei Gao
    \inst{2}
    \and Gang-Yi Zou
    \inst{2}
    \and Wei Wang
    \inst{2}
    \and Zhu-Heng Yao
    \inst{1}
    \and Dong Li
    \inst{3}
    \and Kun Chen
    \inst{3}
    \and Wen Chen
    \inst{3}
    \and Yong-He Zhang
    \inst{3}
    \and Xu-Hui Han
    \inst{1}
    \and Jing Wang
    \inst{1}
    \and Da-Wei Xu
    \inst{1}
    \and Jesse T. Palmerio\orcid{0000-0002-9408-1563}
    \inst{5}
    \and  Susanna. D. Vergani\orcid{0000-0001-9398-4907}
    \inst{4}
     \and Jian-Yan Wei
      \inst{1}   
   \and Bertrand Cordier
      \inst{5}
   }

   \institute{National Astronomical Observatories, Chinese Academy of Sciences,
             Beijing 100101, China; \textit{qiuyl@nao.cas.cn, xlp@nao.cas.cn,  jsdeng@nao.cas.cn}\\
        \and
         Xi’an Institute of Optics and Precision Mechanics, Chinese Academy of Sciences, Xi’an 710119, China; \\        
        \and          
        Innovation Academy for Microsatellites, Chinese Academy of Science, Shanghai 201203, China;\\  
        \and
        LUX, Observatoire de Paris, Université PSL, Sorbonne Université, CNRS, 92190 Meudon, France; \\ 
         \and 
             CEA/Paris-Saclay, Irfu/Département d'Astrophysique, 91191 Gif-sur-Yvette, France;\\
        \and
            School of Astronomy and Space Science, University of Chinese Academy of Sciences, Beijing 101408, China\\
\vs\no
   {\small Received 202x month day; accepted 202x month day}}

\abstract{ 
The 44-cm Visible Telescope (VT)  aboard the Space-based Variable Objects Monitor (SVOM) is a dual-band (400–650 nm and 650–1000 nm) instrument designed to detect and characterize the optical counterparts of gamma-ray bursts (GRBs) and other high-energy transients. This paper presents the VT's design, scientific objectives, observing strategies, and  both space- and ground-based data processing pipelines, along with its first-year in-orbit performance. In-orbit commissioning tests confirm a sensitivity of 22.5 AB mag (300 s exposure), extendable to $\sim\!24$ AB mag through stacking. This performance  enables the VT to monitor over 100 GRBs in its first year with an exceptional $\sim\!80\%$ detection rate for \textit{SVOM}/ECLAIRS-triggered bursts and ToO-observed bursts from other missions (e.g., \textit{Swift, Fermi, Einstein Probe (EP)}), outperforming \textit{Swift}/UVOT’s $\sim\!40\%$ detection rate. 
Beyond its exceptional detection efficiency, the VT played a key role in identifying high-redshift GRBs—most notably GRB 250314A (z = 7.3). Its deep upper limits at long wavelengths (up to 1 $\mu$m) were pivotal in guiding follow-up observations with large ground-based telescopes, enabling crucial near-infrared (NIR) detections. With its rapid response, deep sensitivity, and real-time processing capabilities, the VT is a key instrument for GRB  research in \textit{SVOM}-era, enabling critical studies of GRB optical afterglows, circumburst environments, relativistic jet dynamics, and the origins of optically dark bursts.
}
   \authorrunning{Y. Qiu, L. Xin, J. Deng, et al. }            
   \titlerunning{\textit{SVOM}/VT: Instrument Overview and First-Year Performance}  

   \maketitle
   
\section{Introduction}           
\label{sect:intro}

The study of GRBs -- among the most energetic cosmic transients -- has advanced significantly over the past three decades. Their cosmological origins were firmly established through redshift measurements of their optical counterparts, beginning with the discovery of GRB 970228\citep{1997Natur.387..783C, 1997Natur.386..686V}. It is now widely accepted that long-duration (Type II) GRBs arise from the core collapse of massive stars. Such progenitor models have been substantiated by the observed association of specific nearby GRBs with optical supernovae, most notably GRB 980425/SN 1998bw \citep{1998Natur.395..670G} and GRB 030329/SN 2003dh \citep{2003Natur.423..847H}. Furthermore, the detection of GRB 170817A coincident with a gravitational wave event and a subsequent kilonova counterpart confirmed the long-hypothesized origin of short-duration (Type I) GRBs as mergers of compact objects \citep{2017ApJ...850L..40A}. 

Substantial progress in our understanding of GRBs has been driven by the \textit{Swift} mission \citep{2004ApJ...611.1005G}, a dedicated satellite launched in late 2004 that has remained operational for over two decades. Equipped with a multi-wavelength instrument suite, \textit{Swift} has discovered more than 1,000 GRBs to date, enabling redshift measurements for over 300 of them, including GRB 090423, which holds the record for the highest spectroscopic redshift ($z=8.2$; \citealt{2009Natur.461.1254T}). Its X-ray Telescope (XRT) can observe GRBs from very early times -- following their detection and localization by the Burst Alert Telescope (BAT) -- and has discovered complex, ubiquitous light curve features such as flares and plateaus in early X-ray afterglows, revealing delayed activity of the GRB central engine \citep{2006AIPC..836..392Z,2006ApJ...642..389N}. 

Unlike the XRT, which has detected X-ray afterglows in more than 90\% of cases, \textit{Swift}’s UltraViolet/Optical Telescope (UVOT) shows a modest detection rate ($\sim\!40\%$; \citealt{2023Univ....9..113O}) for optical afterglows. This is likely due to the relatively low quantum efficiency (QE) of its intensified CCD and its long-wavelength cut-off at 650 nm. For GRBs with redshifts greater than $\sim 4$, no photons below 650 nm reach the UVOT due to H Ly$\alpha$ blanketing. Moreover, some optical afterglows may suffer heavily from dust extinction and reddening \citep{2009AJ....138.1690P,2009RAA.....9.1103Z}, rendering them too faint to be detected in the UVOT bands.

The \textit{SVOM} satellite, a joint Chinese-French mission dedicated to gamma-ray burst (GRB) studies \citep{2015arXiv151203323C,2016arXiv161006892W}, adopts a payload architecture similar to \textit{Swift}, combining wide-field instruments for GRB detection and localization with narrow-field telescopes for rapid follow-up observations of lower-energy photons. However, \textit{SVOM}  introduces unique enhancements to overcome \textit{Swift}’s detection limitations, featuring four core instruments: ECLAIRS (a coded-mask gamma-ray imager, \citet{Godet+etal+2026}), the Gamma-Ray Monitor (GRM) (non-imaging, for broadband spectral coverage, \cite{Sun+etal+2026} ), the Micro-channel X-ray Telescope (MXT) (for high-resolution soft X-ray follow-up, \cite{gotz+etal+2026}), and the VT (for optical/NIR afterglow studies). Supported by dedicated ground-based telescopes, this suite is optimized for better detection and spectroscopic confirmation of high-redshift GRBs, characterization of GRB afterglows, and coordination of multi-messenger campaigns.

Specifically, the 44-cm aperture VT \citep{2020ApOpt..59.3049F} is designed to cover the 400-650 nm and 650-1000 nm bands simultaneously, achieving deep detection limits ($3\sigma$) of 22.5 mag (5-min integration) and $\sim\!24$ mag (up to 1.5-hr integration). Consequently, it is capable of detecting GRB optical counterparts up to redshift of $z\sim\!7$ and enabling a quick strategy using optical colors to pre-select high-redshift candidates \citep{2020RAA....20..124W} for timely infrared spectroscopic follow-up with large ground-based telescopes. It can provide observations for   90\% of the satellite-slewed \textit{SVOM} GRBs during their early phase, compared with $\lesssim 13\%$ for any single ground-based telescope. Furthermore, it can conduct ToO observations for non-slew events and GRBs localized by other missions like \textit{Swift} and \textit{EP} \citep{2025SCPMA..6839501Y}.

In this paper, we present an overview of the \textit{SVOM}'s VT payload, covering its scientific objectives (Section \ref{sect:scien}), instrument design (Section \ref{sect:inst}), observing programs and strategies (Section \ref{sect:observ}), and data processing pipelines (Section \ref{sect:pipe}). We also evaluate its on-orbit performance (Section \ref{sect:perf}) and discuss first-year scientific results (Section \ref{sect:resu}), concluding with key findings and future perspectives (Section \ref{sect:summ}).

\section{Scientific objectives }
\label{sect:scien}

The primary scientific objective of the VT is to detect and precisely localize (with sub-arcsec accuracy) the optical counterparts of \textit{SVOM}-triggered GRBs, as well as to characterize their optical afterglows. A key focus is the detection and study of high-redshift GRBs ($z>6$). Additionally, the VT will investigate the nature of optically dark GRBs using a uniform observational sample. Further details are provided below:

\subsection{Rapid optical counterpart localizations for \textit{SVOM} bursts}

Accurate and prompt localization of GRB afterglows in the optical/near-infrared is crucial for directing large ground-based telescopes to measure their redshift while the afterglow remains bright -- a key parameter for understanding these extreme cosmic explosions \citep{2009ARA&A..47..567G,2018pgrb.book.....Z}. To achieve this, the VT initiates follow-up observations immediately after spacecraft stabilization post-slew. The data are then rapidly downlinked to the ground to enable prompt identification and precise astrometric measurements.

However, minimizing latency in downlinking comprehensive data poses a challenge. To address this, \textit{SVOM} employs a real-time VHF channel \citep{Cordier+etal+2026b} to transmit quick-look results. Due to the VHF channel’s limited data rate, onboard processing is required to reduce data volume before transmission. Specifically, \textit{SVOM} uses a dedicated onboard processing unit to analyze early VT images, transmitting the processed results via VHF shortly after GRB detection. Further details on this pipeline are discussed in Section \ref{sect:vopp}.

\begin{figure*}
   \centering
 \includegraphics[width=0.8\linewidth]{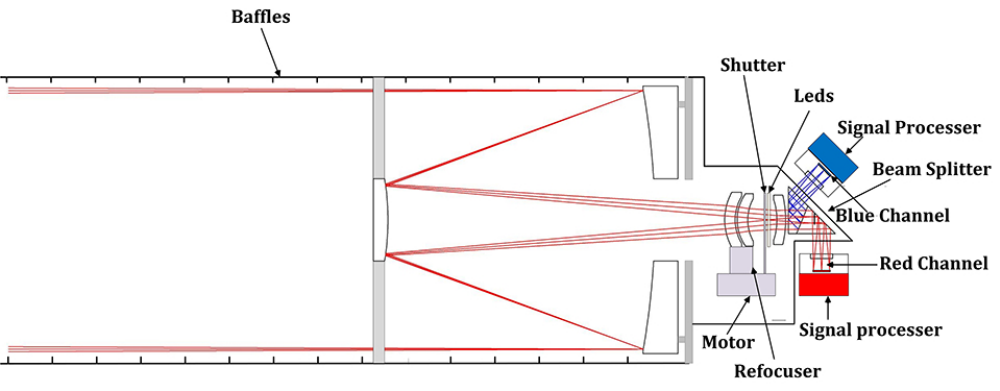}  
	\caption{Schematic of the VT optical design, showing the primary/secondary mirrors, dichroic beam splitter, baffles, and dual-channel CCD detectors (blue: 400–650 nm; red: 650–1000 nm). \vspace{8mm} }
    \label{fig:VT_scheme}
\end{figure*}

\subsection{Characterization of GRBs in the optical band}

Multiband observations of GRBs are a key scientific objective of the \textit{SVOM} mission. Leveraging the high sensitivity and simultaneous dual-band capabilities of VT, \textit{SVOM} can monitor GRB afterglows down to very faint magnitudes ($\sim$ 24 mag). These uniform two-band observations will not only track the evolution of GRB afterglows but also reveal fine structures in their light curves. By eliminating inconsistencies arising from heterogeneous passbands or telescope sensitivities, this approach enables clearer identification of features such as flares, jet breaks, and supernova/kilonova signatures, providing valuable insights into the GRB circumstellar environment and progenitor systems \citep{2016ApJ...817....8P}.

\subsection{Optically dark bursts}

The nature of ``dark bursts'' -- GRBs without detected optical afterglows -- remains unresolved. \citet{2004ApJ...617L..21J} defined them as GRBs with no optical counterpart detected down to R = 21-22 mag within one hour after the burst. Early and deep follow-up observations are crucial for accurately determining the true fraction of dark GRBs. For example, the GROND telescope achieved an impressive $\sim\!80\%$ detection rate within 30 minutes of the trigger, even though it observed only 13\% of cases due to ground-based observational constraints \citep{2024A&A...691A.158G}. Systematic multi-band studies are essential to distinguish among intrinsic faintness, high dust extinction \citep{2009AJ....138.1690P,2009RAA.....9.1103Z}, and extreme redshifts \citep{2009Natur.461.1254T} as possible origins. The space-borne dual-band VT is expected to help resolve this decades-old puzzle, benefiting from deep optical coverage extending to 1 $\mu$m and the ability to observe more than 90\% of \textit{SVOM} GRBs without interruptions from weather or day-night cycles.

\subsection{High-redshift GRBs }

The detection and identification of high-redshift gamma-ray bursts (GRBs) at $z > 6$ is a high-priority core science objective of the \textit{SVOM} mission.
ECLAIRS is specifically optimized for detecting high-z GRBs, owing to its enhanced sensitivity in the low-energy X-ray range. Complementing this capability, the VT extends the wavelength coverage of optical observations up to 1 $\mu$m, enabling the identification of optical counterparts up to $z\sim 7$. Additionally, the VT’s deep early-phase exposures, color analyses, and stringent upper limits in cases of non-detection play a crucial role in the rapid pre-selection of high-redshift candidates \citep{2020RAA....20..124W}. A notable example is GRB 250314A at $z=7.3$ \citep{2025arXiv250718783C}, which highlights the VT’s pivotal contribution to high-redshift GRB science in the \textit{SVOM} era.

\section{Instrument design and capabilities}
\label{sect:inst}

The optical design, detector system, and operational architecture of the VT are carefully balanced to maximize GRB detection efficiency.   Detailed technical specifications and ground test results—including Qualification Model validation—are reported in prior studies \citep{2020ApOpt..59.3049F,2020SPIE11443E..0QF,2021Photo...8..132P,2022PASP..134c7001P}, with Flight Model performance documented in \citet{Zhang+etal+2026}.  

This section outlines the instrument's core subsystems (see VT schematic in Fig. \ref{fig:VT_scheme}), including: optics, beam splitter, detector, calibration unit, baffle system, and focusing mechanism.

\subsection{Optics}

The VT employs a 44-cm aperture Ritchey-Chrétien optical system with an $f/8$ focal ratio. Its primary mirror, constructed from a lightweight and rigid SiC, is coated with high-reflectivity silver to ensure optimal performance across the operational wavelength range. The design incorporates an optimized secondary mirror to accommodate a Fine Guidance Sensor (FGS) on the focal plane, maximizing the field of view (FOV) while minimizing throughput losses. This configuration preserves full system functionality and performance.

\subsection{Beam splitter}

\begin{figure}
    \centering

        \includegraphics[width=0.8\textwidth]{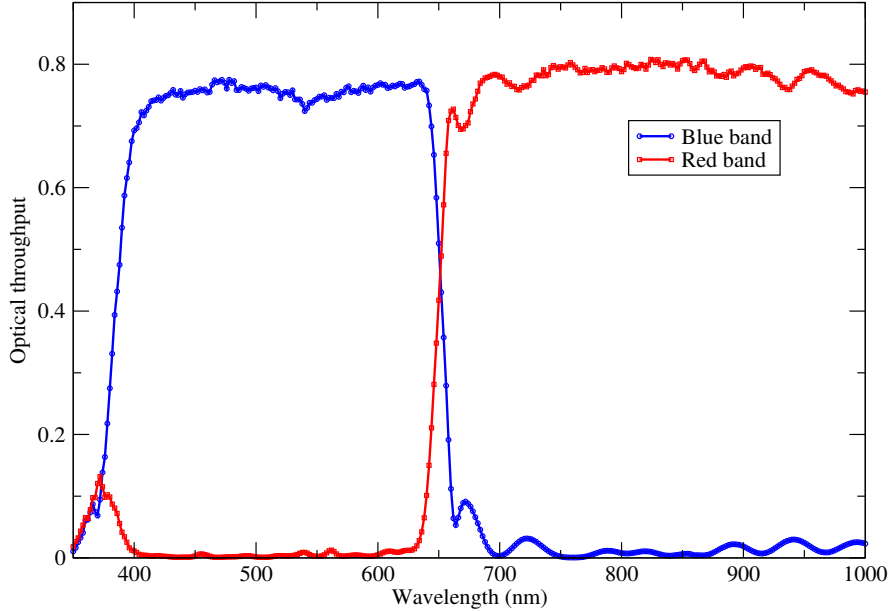}
    \caption{Ground-calibrated throughput of the VT system for blue and red channels (CCD QE excluded).}
    \label{fig:VT_transmission_curves}
\end{figure}

To optimize observational efficiency and minimize mechanical risks associated with filter wheels, the VT utilizes a dichroic beam-splitter configuration for simultaneous dual-band observations (blue: 400–650 nm; red: 650–1000 nm). This design facilitates high-precision analysis of GRB afterglow chromatic evolution, real-time transient classification, and rapid data acquisition. The dichroic system exhibits $>95\%$ efficiency in both spectral bands, ensuring precise co-alignment of the fields of view and maximizing observational temporal resolution and throughput for rapidly evolving transients such as GRBs. The ground-calibrated throughput profiles of the VT’s flight model, depicted in Figure \ref{fig:VT_transmission_curves}, account for mirror reflectivity, obscuration effects (primarily by the secondary mirror), and dichroic performance, but do not include CCD quantum efficiency (QE).

\subsection{Detectors}
\begin{figure}
    \centering
    \includegraphics[width=0.8\linewidth]{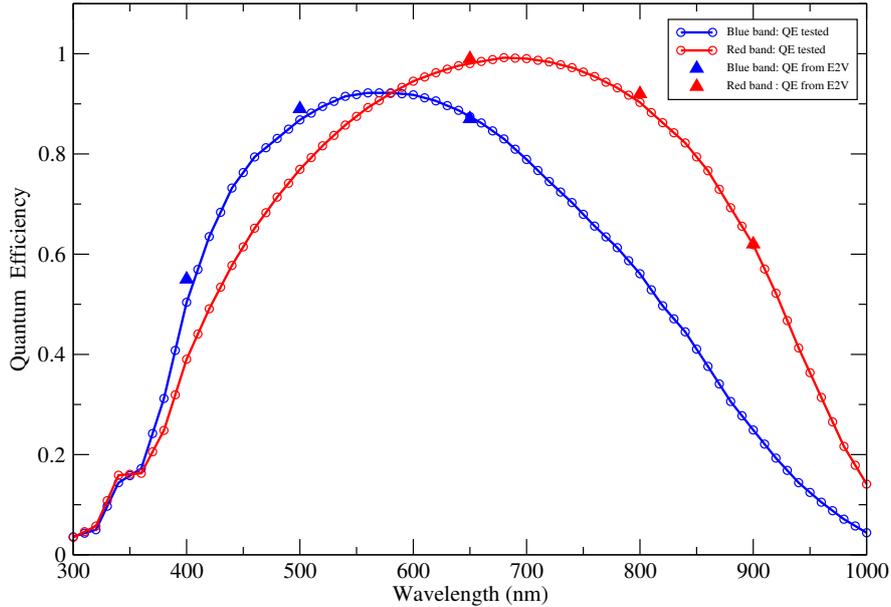}
    \caption{QE curves for the VT’s blue and red CCDs, measured pre-launch. The red channel’s deep-depletion CCD shows enhanced sensitivity at $\lambda > 900$ nm.}
    \label{fig:VT_detector_QE}
\end{figure}

 Each of the two VT channels is equipped with a 2048 $\times$ 2048 e2v CCD detector (13.5 $\mu$m pixel size), optimized for its respective wavelength range. The blue channel employs a thinned, back-illuminated CCD with a broadband coating, achieving a peak QE of $> 90\%$ at 550 nm, while the red channel uses a deep-depletion CCD with enhanced response at longer wavelengths (QE $> 60\%$ at 900 nm compared to $\sim$20\% for a standard CCD), as displayed in Figure \ref{fig:VT_detector_QE}. They share the same pixel scale of 0.764 arcsec and a 26 $\times$ 26 arcmin FOV, which is well matched to the typical localization error box of ECLAIRS detections for efficient follow-up.

The CCDs feature a frame-transfer architecture to eliminate the need for long-life, high-reliability mechanical shutters and minimize readout dead time. However, the absence of a shutter allows photon leakage during the frame transfer operation, introducing smear effects. In particular, bright stars can produce prominent vertical streaks that inate nearby faint sources. Nevertheless, this design trade-off was accepted to avoid the risk of mechanical shutter failure.

The detectors operate at $-65^\circ$C (blue channel) and $-75^\circ$C (red channel), with a temperature stability of $\pm$0.5°C. 
Dark currents are low at these temperatures (0.1$\,e^{-}/\text{pix}/s$ for the blue channel and 0.3$\,e^{-}/\text{pix}/s$ for the red channel), making the sky background, rather than detector noise, the dominant noise source in VT images.


\subsection{Calibration unit}

A calibration unit is integrated into the VT to acquire bias, dark, and flat-field frames for the correction of instrumental effects in VT images. A motorized shutter—coated with a diffuser and positioned near the beam splitter—functions as a flat-field calibration source when illuminated by integrated multi-wavelength LEDs. Each channel offers three selectable wavelengths: 470 nm, 527 nm, and 640 nm (blue); and 670 nm, 740 nm, and 870 nm (red). With the shutter closed, the system captures bias frames (zero-second exposures), dark current frames (LEDs off), and multi-wavelength flat-field frames (LEDs on). These calibration data are used for both onboard observation processing and ground-based data reduction.


\subsection{Stray light suppression and baffles }

The VT employs a multi-layered stray light suppression strategy, incorporating anti-reflective coatings on the internal barrel, light-trapping vanes, and a baffle system that includes both internal baffles (on the primary and secondary mirrors) and an external baffle at the entrance. While these measures enhance optical performance, the length of the external baffle is constrained to preserve the unobstructed FOV of ECLAIRS. This limitation degrades stray-light suppression, permitting Earthshine contamination. Observations are thus confined to Earth's umbra ($\sim35$ min/orbit), but the reduced observing time remains acceptable for continuous GRB monitoring.

\subsection{Focusing mechanism}

The corrector lens group, located upstream of the beam splitter, performs two key functions: compensating for field aberrations and fine-tuning focus. A motorized translation stage controls the axial displacement of the lens group, allowing for precise and dynamic focus adjustment.

\section{Observing programs and strategies}
\label{sect:observ}

While the comprehensive observing programs and strategies of the \textit{SVOM} mission are detailed in \citet{Cordier+etal+2026a}, this section focuses on the observing strategies specific to the VT for each program.

\subsection{GRB follow-up observations}

When ECLAIRS detects a GRB and the signal-to-noise ratio (SNR) exceeds a predefined threshold, the platform automatically slews toward the burst source. Prior to full platform stabilization, the VT acquires 15-second ``chance mode'' exposures to mitigate the effects of jitter. Upon achieving stability, the VT switches to longer exposures (configurable, with a nominal duration of 50 seconds). High-quality observations are typically achievable for 30–40 minutes per 96-minute orbit while in the Earth's umbra. Outside this period, Earthshine significantly degrades the sensitivity to faint objects, thereby limiting data quality.

During the first two orbits, the Payload Data Processing Unit (PDPU) processes four observation sequences onboard. The resulting data products are transmitted via the VHF data link to the French and Chinese Science Centers. Following astrometric and photometric calibrations, sources extracted and photometrically measured by the onboard pipeline (VOPP; see Section \ref{sect:vopp}) are cross-matched with archived catalogs to identify GRB candidates. VT instrument scientists, in collaboration with Burst Advocates (BAs), then analyze these candidates alongside 1-bit sub-images to confirm bright optical counterparts.

All raw images are stored in the onboard mass memory and downlinked to the ground via X-band antennas. These images are processed at the VT Instrument Center using the VXPP pipeline (see Section \ref{sect:vxpp}). Leveraging the complete dataset, the pipeline searches for fainter optical counterparts of GRBs that may have been undetected in the real-time analysis. Once a GRB is confirmed, the VXPP pipeline automatically performs photometry on both blue- and red-band images.

Following the automatic platform slew, the VT observes GRBs over five consecutive orbits. For the first two orbits, it operates in full-frame mode. If the MXT payload successfully localizes the GRB during this period, the VT transitions to window mode, centered on the MXT-derived position, from the third orbit onward. If no localization is achieved, the VT defaults to interval mode, reducing data volume by storing only a subset of frames (e.g., every second or third frame). This strategy optimizes memory usage while ensuring critical data retention. 

\subsection{ToO observations}
\label{sect:ToO}

If a GRB detected by ECLAIRS falls below the slew threshold, no autonomous platform slew is executed. Instead, follow-up observations with VT (and MXT) are conducted in ToO mode via ground-uplinked telecommands. During ToO observations, the VT employs long exposures (typically $>50$ s). The choice between window and full-frame readout modes is determined by whether the GRB has been optically confirmed. If a counterpart is confirmed, window mode is typically utilized to minimize data volume; otherwise, the instrument is set to full-frame mode.

Via the BeiDou Navigation Satellite System, telecommands for ToO observations can be uploaded to \textit{SVOM} with an average latency of about 20 minutes. The duration of ToO observations is determined by the observation epoch: early-time follow-up typically requires fewer orbits (1–2), whereas late-time observations generally demand more (3–5). 

The software tools supporting ToO scheduling are detailed in \citet{Han+etal+2026}. The above strategies also apply to ToO observations of GRBs and similar transients detected by other missions, such as \textit{Swift} and \textit{EP}. 

\subsection{General Program observations}

Beyond GRB-triggered observations and ToO programs, \textit{SVOM} executes a General Program (GP) for pre-scheduled targets, such as AGNs, X-ray binaries, flare stars, and other variable objects. These observations, typically performed simultaneously with the MXT, account for approximately half of the VT’s total observing time. The GP is open to \textit{SVOM} Co-Investigators (Co-Is), who are required to submit proposals at least six months in advance. Approved observations are scheduled on a monthly or yearly basis by the \textit{SVOM} Mission Center. Details regarding the GP proposal and scheduling processes are presented in \citet{Han+etal+2026}.

\begin{table}[]
    \centering
    \caption{Configuration Parameters for VT ToO and GP Observations}
    \begin{tabular}{c|l}
\hline
Parameters & Notes \\
\hline
Exposure Time & Optimize the exposure time\\
              & by balancing the signal-to-noise \\
              & ratio (SNR) with temporal\\ 
              & resolution while preventing\\ & signal saturation\\
\hline
Window Size   & Ensure the window size is \\                & large enough to account for\\               & platform pointing accuracy \\
              & while minimizing data volume\\
\hline
Read channel  & There are three options:\\                  & LEFT, RIGHT or BOTH. \\
              & The BOTH is proposed \\
              & when the full frame mode \\
              & and short exposure is used\\ 
\hline
Read speed  & Range: from 75 Hz(low-noise) \\
            & to 200 Hz (fast readout) \\
              & Trade-off: Higher speeds\\
              & reduce readout time but\\ &increase readout noise.\\
\hline
Frame Transfer Mode & Enabled: Continuous exposures \\
            & Disabled: Next exposure begins\\ 
            & only after full readout of the\\
            & prior frame\\
                            
\hline
Readout Mode       & Continuous Mode: Save all images \\
                   & Interval Mode: Save  images \\
                   & with an interval of 1 or 2 frames \\
\hline

    \end{tabular}
    
    \label{tab:VT_configuration}
\end{table}

For both ToO and GP observations, the VT requires six configuration parameters (Table \ref{tab:VT_configuration}): exposure time, readout window size, readout channel (LEFT, RIGHT, or BOTH), readout speed (75–200 Hz), Frame Transfer Mode, and readout mode (continuous or interval-based storage).
The VT Instrument Center provides two tools (\url{http://159.226.170.70/cgi-bin/calsnr1.py}) for GP and ToO proposers: a photometric system converter (for converting magnitudes between the VT and other systems) and an Exposure Time Calculator (ETC) to estimate the required integration time.
\subsection{Calibrations}

\subsubsection{Internal calibration}

The instrument supports two dedicated calibration modes: one for onboard processing and another for ground data processing. 

\begin{enumerate}
    \item \textbf{Onboard Processing} – Due to onboard memory constraints  and to simplify data processing, we employ dark frames (with exposure times matching GRB observations) to correct for both dark current and CCD bias \citep{Cai+etal+2026}.  Flat-field images are acquired independently for each CCD using LEDs at 527 nm (blue) and 740 nm (red), while dark frames are captured simultaneously for both CCDs. For each CCD, master flat and dark images are generated by combining six individual frames (with cosmic-ray rejection) to improve SNR. All four master calibration images (two flats, two darks) are stored in the PDPU for real-time processing via the VOPP.

    \item \textbf{Ground Processing} – Captures flat fields across all LED wavelengths, along with long-exposure and multiple dark frames. All data are downlinked for comprehensive spectral response analysis and ground-based calibration.
    
\end{enumerate}

\subsubsection{External calibration}

External calibration involves using standard stars to calibrate the instrument. Spectrophotometric standards, typically highly stable white dwarf stars with precise fluxes, are employed to determine the magnitude zero point, which is tied to the telescope’s throughput. Since the VT is subject to post-launch contamination, these standards also help monitor the evolution of contamination effects \citep{Yao+etal+2026}. Additionally, we plan to observe photometric standards to verify the spectral response shape.

\section{Pipelines for VT data processing}
\label{sect:pipe}

Several pipelines exist for VT data processing, spanning from onboard to ground operations. Below, we outline their workflow and processing strategy.

\subsection{VT Onboard data Processing Pipeline (VOPP)}
\label{sect:vopp}

The VHF data-link enables near real-time transmission of burst data from the spacecraft to the ground. However, due to the large size of VT images (2k × 2k pixels), direct transmission is impractical, requiring efficient onboard data processing to reduce the data volume. To address this, we employ a selective processing strategy, focusing only on high-quality images to create four observation sequences during the first two post-trigger orbits. Further details on this approach can be found in \citet{2022SPIE12181E..5SM}. Below, we provide a brief overview of the VOPP workflow and its resulting data products; additional details are available in \citet{Cai+etal+2026}.

The VOPP workflow comprises four key steps to ensure efficient data reduction while maintaining scientific integrity. First, an image quality assessment verifies platform stability and background levels. Next, instrument calibration applies bias, dark, and LED flat-field corrections (executed on the FPGA) to refine the data. The third step, image stacking, improves the signal-to-noise ratio (SNR) while mitigating cosmic-ray contamination. Finally, source extraction and aperture photometry are performed to quantify light intensity. This structured pipeline enhances data reliability and accuracy at every stage.

The pipeline generates three key products: Attitude Charts (ATCs), Finding Charts (FDCs), and 1-bit images. Details for each are presented below.

\subsubsection{ATC}

An ATC contains pixel coordinates for 21 bright, unsaturated stars in observed images. These coordinates are quickly downlinked via VHF and matched with their J2000 reference positions from a known catalog (e.g., \textit{Gaia}) using ground-based software. This process enables sub-arcsec level pointing attitude determination. Details on the fast star extraction algorithm can be found in \citet{2010SCPMA..53S..51W}. The ATC serves two main purposes: performing astrometry for VT source lists and assisting MXT in achieving enhanced localization.

\subsubsection{FDC}

An FDC contains a source list above a predefined SNR threshold, extracted from combined sub-images centered on the MXT localization. It covers the MXT error box, with cosmic ray contamination largely removed through the combination of 3–6 images. For each detected source, the FDC provides three aperture photometry measurements: Small (optimized for compact cores), Intermediate (balanced sensitivity for point sources), and Large (i.e., isophotal, for morphological analysis), along with ellipticity data. By analyzing these photometric measurements, blended sources and artifacts can be identified and flagged in the on-ground pipelines, thereby reducing the number of false candidates.

\subsubsection{1-bit image}

A 1-bit image is a digitized composite of sub-images in which each pixel is assigned a binary value (0 or 1) depending on whether its flux exceeds the detection threshold. This format allows scientists to efficiently distinguish genuine astronomical sources from image artifacts through visual inspection. After compression using Run-Length Encoding (RLE), these 1-bit images are nearly two orders of magnitude smaller than their original 16-bit counterparts. These compressed images, combined with FDCs, facilitate the rapid early identification of bright optical counterparts during observations of a GRB.

\subsection{VT VHF data Processing Pipeline (VVPP) and VT Afterglow Candidates  Pipeline(VTAC)}

Detailed descriptions  the two ground pipelines, VVPP and VTAC, are provided in \citet{wu+etal+2026}. Here, we briefly outline their purposes and functionalities.

\subsubsection{VVPP}
\label{sect:vvpp}
This pipeline conducts both astrometric and photometric calibrations for VOPP products, focusing on the source list (FDC). For astrometry, the Gaia catalog serves as the reference. Initial calibration is applied to the ATC because it is received earlier; however, due to the limited number of stars (21 across the FOV), the resulting accuracy is limited to 0.1–0.2 arcsec. Refined astrometry using the FDC, which incorporates a significantly larger star sample, improves the astrometric accuracy to ~0.02 arcsec. For photometry, aperture corrections are applied, and instrumental magnitudes are calibrated to VT AB magnitudes using two band-specific zero-points.

\subsubsection{VTAC}

This pipeline identifies optical counterparts of GRBs by cross-matching VVPP products with known catalogs and analyzing preliminary light curves derived from four observing sequences spanning two orbits. It automatically generates bright candidates, which can be reported to the GRB Coordinate Network (GCN). Additionally, the pipeline includes a visualization tool for comparing sources in the FDC with existing catalogs, aiding scientists in identifying fainter candidates more efficiently.

\subsection{VT X-band data Processing Pipelines (VXPP)}
\label{sect:vxpp}

This pipeline processes VT images downlinked via the X-band antenna. A detailed description of the pipeline is provided in \citet{Li+etal+2026b}. Below, we outline its basic functionalities.

\subsubsection{Reconfirmation of optical counterparts}

Since X-band data provide more comprehensive information than VHF data, reconfirming optical counterparts identified by VVPP and VTAC is essential. To achieve this, we employ a two-step process. First, we use the latest calibration data to verify whether candidate signals originate from bad pixels. Visual inspections are conducted when necessary to distinguish genuine objects from artifacts, such as hot pixels or other detector defects. Second, we perform light curve analysis to further validate the candidate, leveraging the rapid variability characteristic of GRBs to differentiate them from ordinary variable stars.

\subsubsection{Search for new optical counterparts}

If optical counterparts remain unidentified by the VVPP and VTAC using VHF data, the next step involves searching for counterparts via deeper detections achieved through image stacking. However, this approach requires cross-matching with deep catalogs, where contamination from the bright field stars and artifacts (e.g., hot pixels) can complicate identification. To mitigate these challenges, precise X-ray afterglow localization is crucial to minimize interference from spurious sources. Analysis indicates that an X-ray localization accuracy of $<1$ arcmin is necessary to ensure that VT’s deep detection capability can effectively uncover faint optical counterparts.

\subsubsection{Photometry of identified optical counterparts}

Once counterparts are identified, the final task is photometry. This involves processing images using standard procedures, including bias, dark, and flat-field corrections, to remove instrumental effects. An image stacking strategy is then implemented, typically involving stacking fewer images in early phases and more in later phases, consistent with the decaying nature of GRB afterglows. To optimize the SNR, a medium-sized aperture is used, with aperture corrections calibrated using nearby bright stars. Additionally, precise astrometry facilitates automated photometric measurements.

\section{VT Performances in orbit}
\label{sect:perf}

During the commissioning phase, we assessed the VT performance, including detector characteristics, point spread function (PSF; characterized by encircled energy), stray light, throughput, platform pointing stability, detection limits, and co-alignment with other instruments.

\subsection{Detector characteristics}
\subsubsection{Bias and overscan}

Since most sources observed by the VT are point sources, bias drift primarily affects the background levels rather than object fluxes, as photometric pipelines subtract the background to determine the net flux. However, in typical two-port readout configurations, bias drift can create left--right imbalances, complicating image processing. Therefore, continuous bias monitoring and overscan correction --using the 10 dedicated overscan columns per port -- are implemented to mitigate these effects.

\begin{figure}
    \centering
        \includegraphics[width=0.8\linewidth]{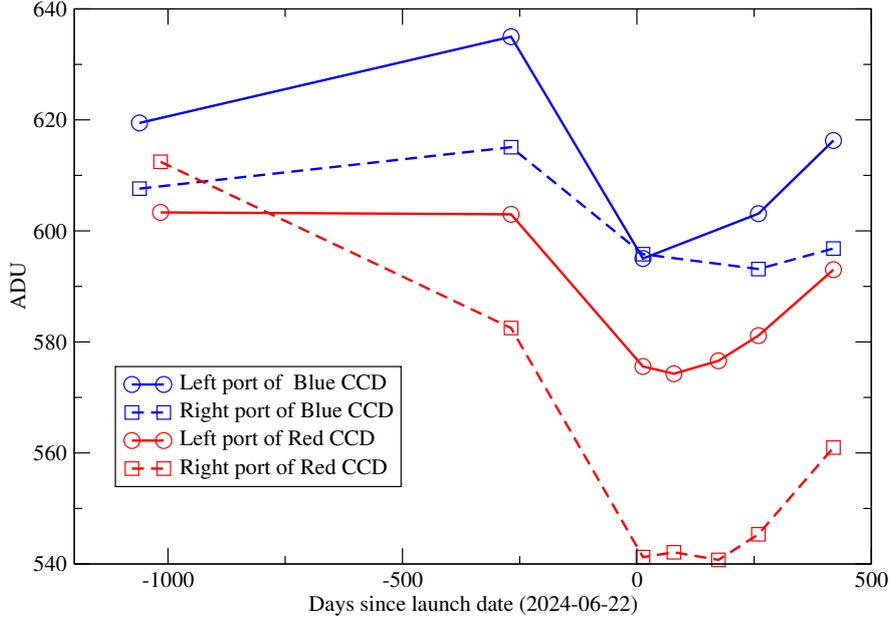}   
        \caption{Pre- and post-launch CCD bias drifts for both blue and red channels.\vspace{12mm}}
    \label{fig:bias_drift}
\end{figure}

\begin{figure}
    \centering
  
  \includegraphics[width=0.8\linewidth]{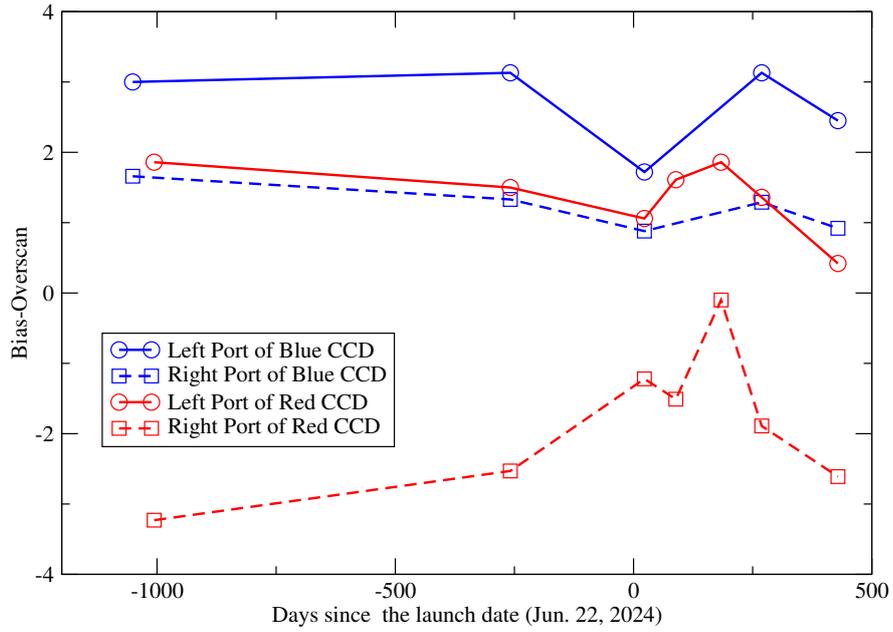}
    \caption{Same as Fig. \ref{fig:bias_drift}, with overscan correction applied.}
    \label{fig:overscan}
\end{figure}

Figure \ref{fig:bias_drift} shows bias drift of several tens of ADU in both channels from 1,000 days before to 400 days after the \textit{SVOM} launch, which cannot be fully corrected by standard bias frames alone. Overscan correction proves effective, reducing the residual drift to within 2 ADU, as demonstrated in Figure \ref{fig:overscan}.

\subsubsection{Hot pixles}

\begin{figure}
    \centering
    \includegraphics[width=0.8\linewidth]{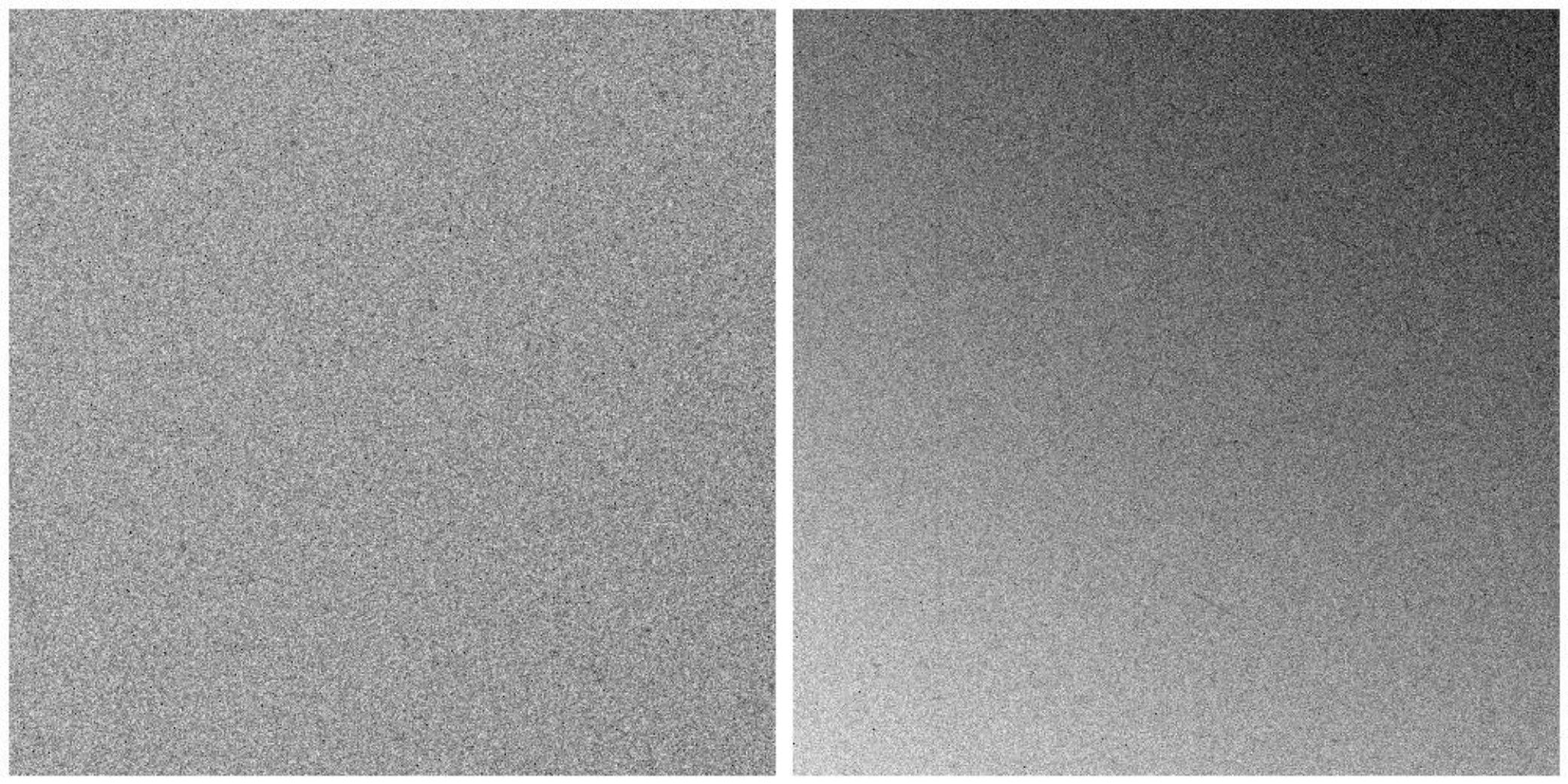}
    \caption{Blue (left) and red (right) channel dark frames (2024 July) show very few hot pixels shortly after launch.}
    \label{fig:dark}
\end{figure}

\begin{figure}
    \centering
    \includegraphics[width=0.8\linewidth]{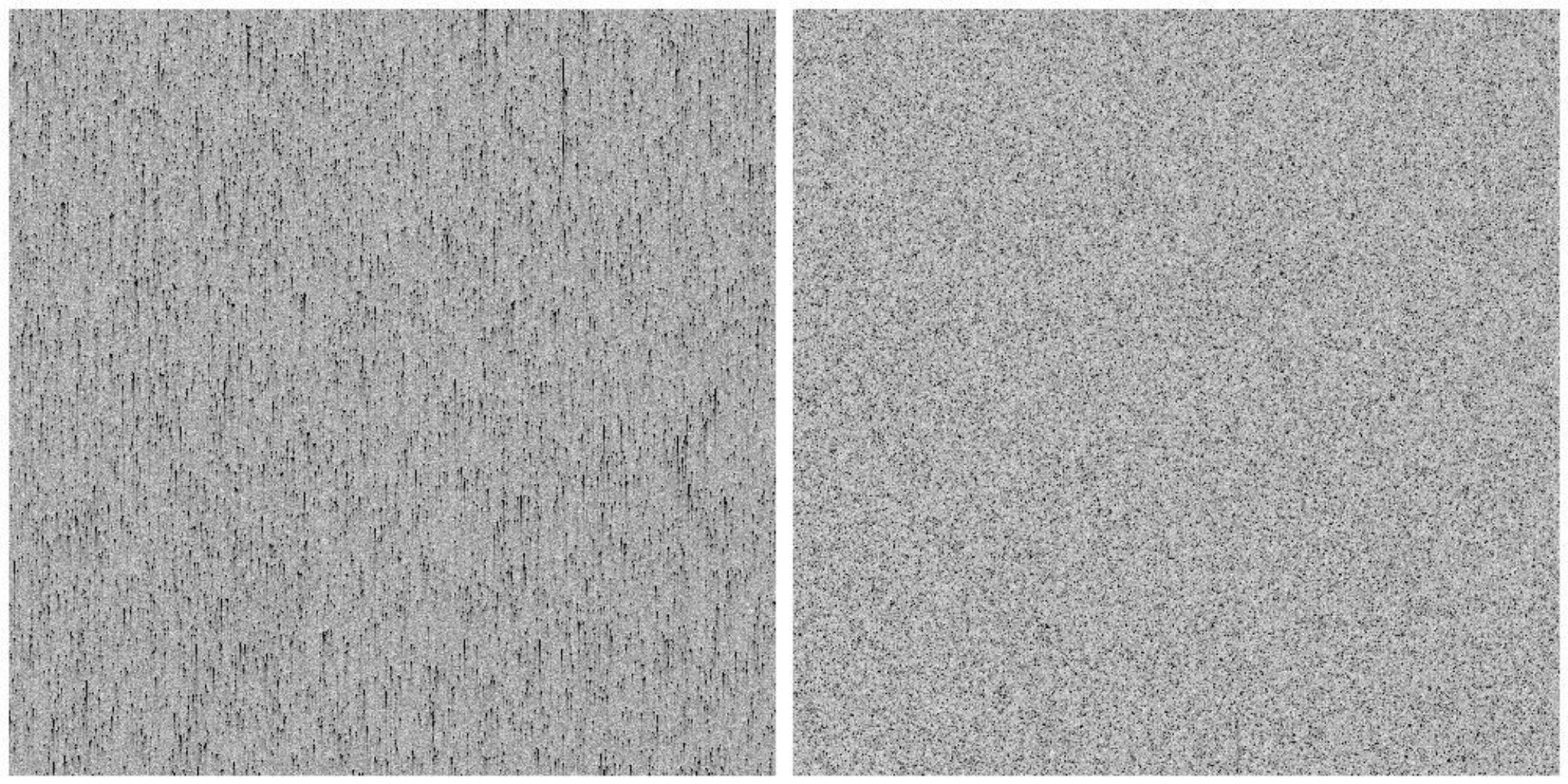}
	\caption{Same as Fig. \ref{fig:dark}, but for August 2025, showing the evolution of hot pixels after one year in orbit. A pronounced contrast is observed between the CCDs: the blue channel exhibits a significant increase in hot pixels, while the red channel remains largely stable, suggesting differential radiation sensitivity.\vspace{16mm}}
    \label{fig:dark_BR_1year}
\end{figure}

Post-launch dark-current monitoring reveals accelerated hot-pixel growth, particularly during periods of high solar activity. Figures \ref{fig:dark} and \ref{fig:dark_BR_1year} illustrate the evolution of hot pixels in dark images over the first year in orbit. The Red channel exhibits better performance than the Blue channel, likely due to two key factors: (1) a lower operating temperature ($-75^\circ$C vs. $-65^\circ$C) and (2) a deep-depletion CCD design that enhances radiation hardness. Nevertheless, detector bake-out proves highly effective in mitigating hot pixels; observations indicate that the hot pixel count is reduced by approximately 50\% following this procedure. A detailed description of the bake-out process will be presented in a future paper \citep{ma+etal+dark+2026}.

One year post-launch, dark frames from the blue channel exhibited hot-pixel trailing (see Fig. \ref{fig:dark_BR_1year} ), consistent with charge transfer inefficiency (CTI) caused by cumulative radiation damage. While CTI produces charge trapping that manifests as trailing in dark frames, analysis of science exposures (including long integrations) showed no detectable trailing or meaningful signal degradation. This operational resilience occurs because the higher background flux in science images provides sufficient charge to fill trapping sites, whereas the low background in dark frames amplifies CTI artifacts. 

\subsubsection{Photon transfer curves}

To construct the photon transfer curves (PTCs), we analyzed pairs of distinct flat-field images with identical flux levels. After removing fixed-pattern noise, bias contributions, and dark current effects, we calculated both the mean flux and its root-mean-square (RMS) fluctuation.

\begin{figure}
    \centering    \includegraphics[width=0.8\linewidth]{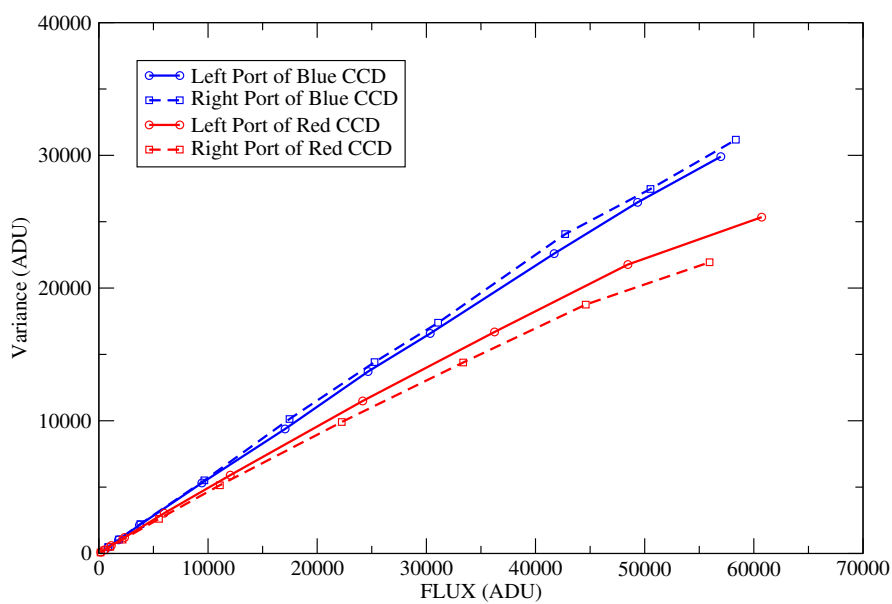}

    \caption{CCD photon transfer curves for both blue and red channels.}
    \label{fig:photon_transfer}
\end{figure}


The resulting PTCs, as shown in Figure \ref{fig:photon_transfer}, demonstrate that the blue channel maintains superior linearity, following the expected inverse gain relationship. In contrast, the red channel exhibits significantly earlier saturation onset. Consequently, the blue channel demonstrates a larger full-well capacity as derived from PTC analysis compared to the red channel. This behavior aligns with previous observations from pre-launch ground testing \citep{2022PASP..134c7001P}.

\subsubsection{Gain Evolution}

\begin{figure}
    \centering
            \includegraphics[width=0.8\linewidth]{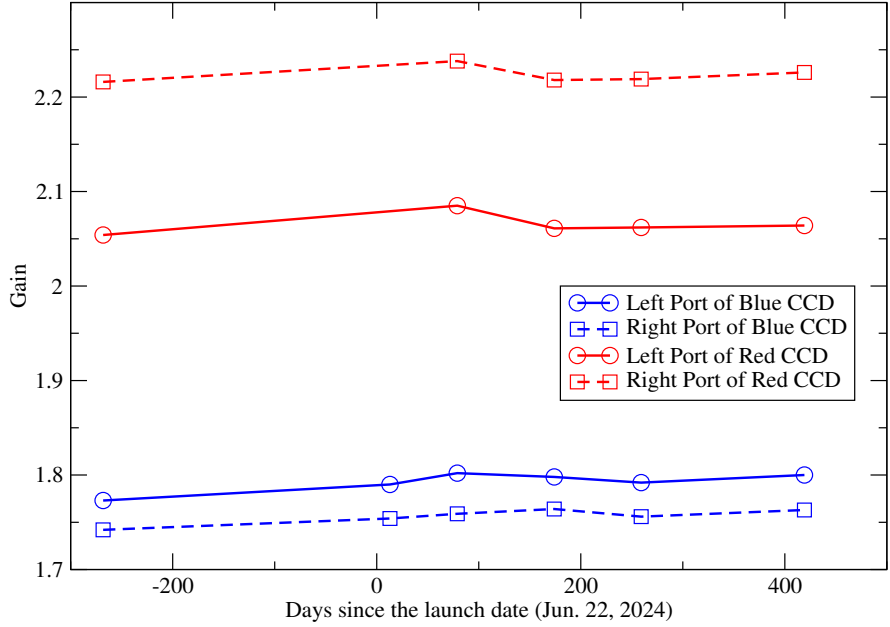}
    \caption{Pre- and post-launch gain evolution for both blue and red channels.}
    \label{fig:GAIN}
\end{figure}

Gain drift can affect the photometric zero-point, making gain monitoring a critical step in data processing. The photon transfer method, the most common and practical approach for determining gain, yields results that depend on the flux level of the flat field used. To ensure consistency, we employ an LED lamp as the flat-field source and maintain fixed exposure times, thereby stabilizing the flux of our flat-field images. 
Figure \ref{fig:GAIN} presents the two-year evolution of detector gains, showing excellent stability in the red channel (±0.5\% variation) while revealing minor but systematic fluctuations in the blue channel (±1.0\%). During data processing, we account for these temporal variations by applying linear interpolation between successive calibration measurements. This correction method suppresses photometric errors to        $<0.1\%$ (red) and $<0.2\%$ (blue), ensuring sub-percent precision across the full observation period.

\subsubsection{Linearity}

The detector linearity is verified in tests using stable light sources, including standard stars (serving as in-orbit natural references) and an adjustable LED lamp, by confirming that the measured flux scales proportionally with exposure time. These measurements form the basis of our calibration procedures.

\begin{figure}
    \centering
        \includegraphics[width=0.8\linewidth]{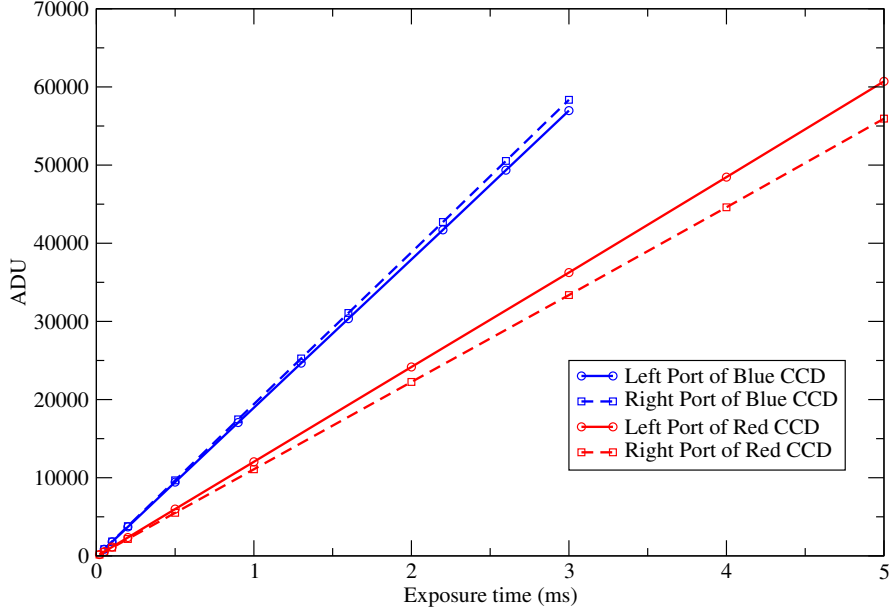}
    \caption{CCD linearity characterization: Mean net flux (ADU) versus LED illumination time (ms) for the blue and red channels.}
    \label{fig:linear}
\end{figure}

As illustrated in Figure \ref{fig:linear}, both detectors exhibit excellent linearity (within 1\%) over varying LED illumination times, with measured flux scaling proportionally up to the 16-bit digitization limit. Minor gain discrepancies between the left and right readout ports were observed but were systematically corrected during flat-field processing to ensure photometric uniformity across the detector array. This test confirms the suitability of the onboard LEDs as linearity calibration sources.

\subsection{Encircled energy (EE)}

Since the launch of \textit{SVOM}, two focus adjustments have been made. The first was conducted after the detector cooled to its nominal temperature, following the initial week-long de-contamination heating phase post-launch. The second adjustment occurred 11 months later to address an observed ~10\% degradation in the EE (energy efficiency), requiring only minor fine-tuning.

\begin{figure}
    \centering
\includegraphics[width=0.7\linewidth]{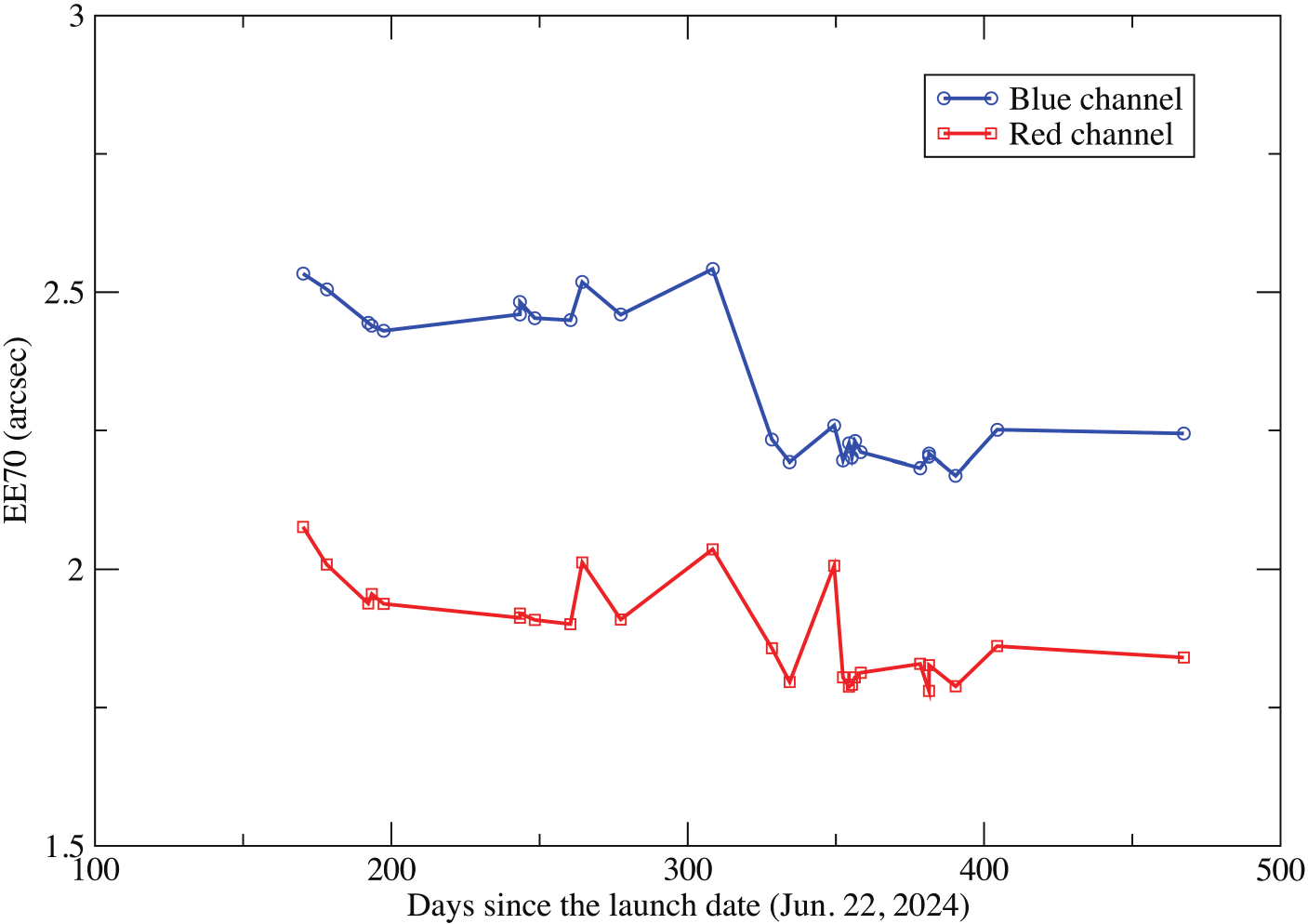}
	\caption{Post-launch EE70s evolution for both blue and red channels.\vspace{8mm}}
    \label{fig:EE70_B_R}
\end{figure}

Figure \ref{fig:EE70_B_R} illustrates the post-launch evolution of the 70\% Encircled Energy (EE70). Following the second adjustment, improvements of 10\% and 5\% were observed in the blue and red channels, respectively. Currently, the blue channel achieves an EE70 of 2.2 arcsec ($\pm{5\%}$), while the red channel reaches 1.8 arcsec ($\pm{5\%}$). This PSF difference is primarily attributed to the deep-depletion CCD used in the red channel, which offers superior Modulation Transfer Function (MTF) and reduced charge diffusion.

Overall, the EE demonstrates excellent long-term stability and short-term consistency in orbit, remaining robust even during large-angle slews (see Section \ref{sect:stability}). These characteristics ensure consistent operation and reliable detection performance for the VT.

\subsection{Stray light}
\label{sect:staylight}

\begin{figure}
    \centering    \includegraphics[width=0.7\linewidth]{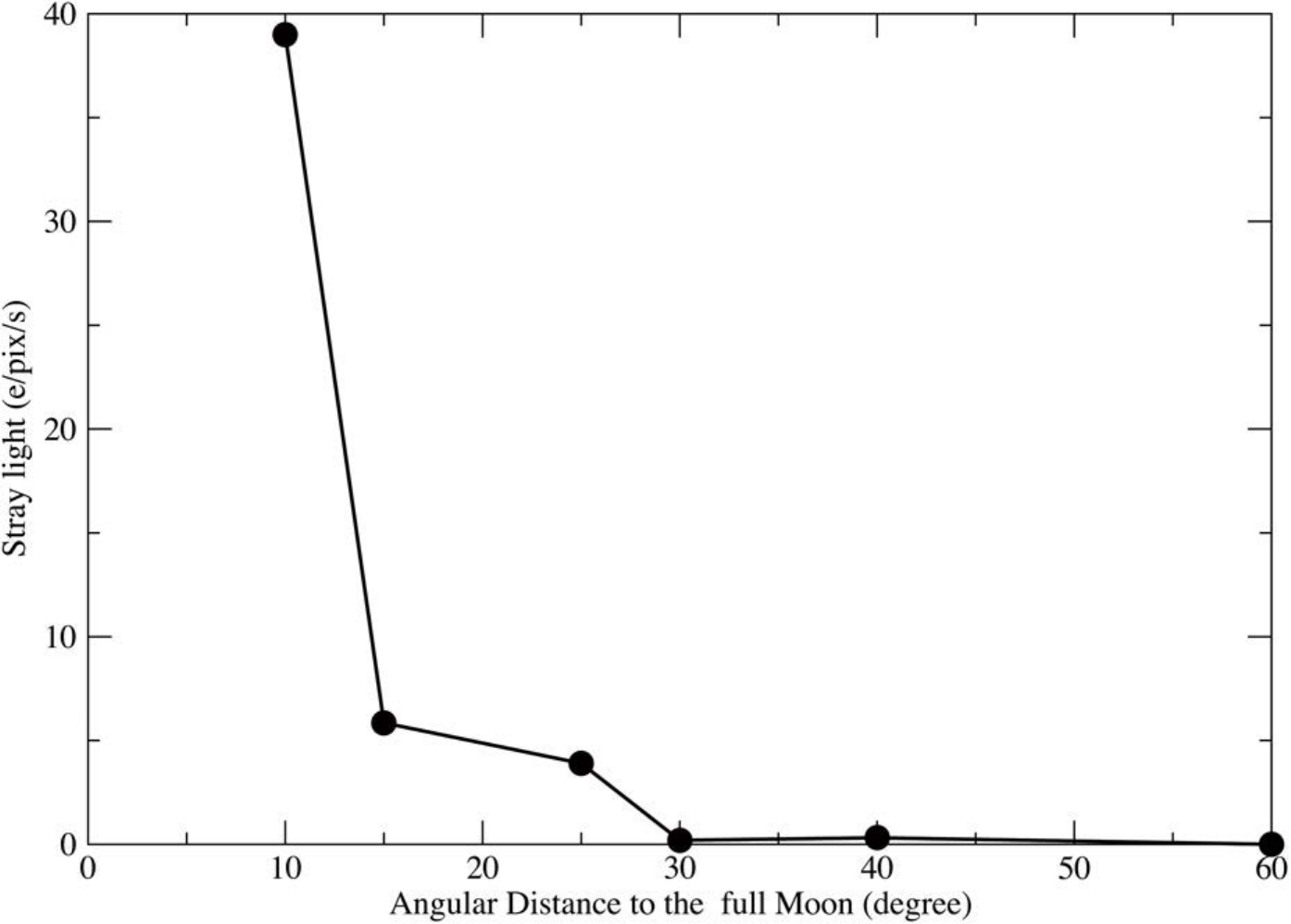}
    \caption{Stray light contamination from the full Moon at multiple off-axis angles.}
    \label{fig:straylight}
\end{figure}

\begin{figure}
    \centering
    \includegraphics[width=0.8\linewidth]{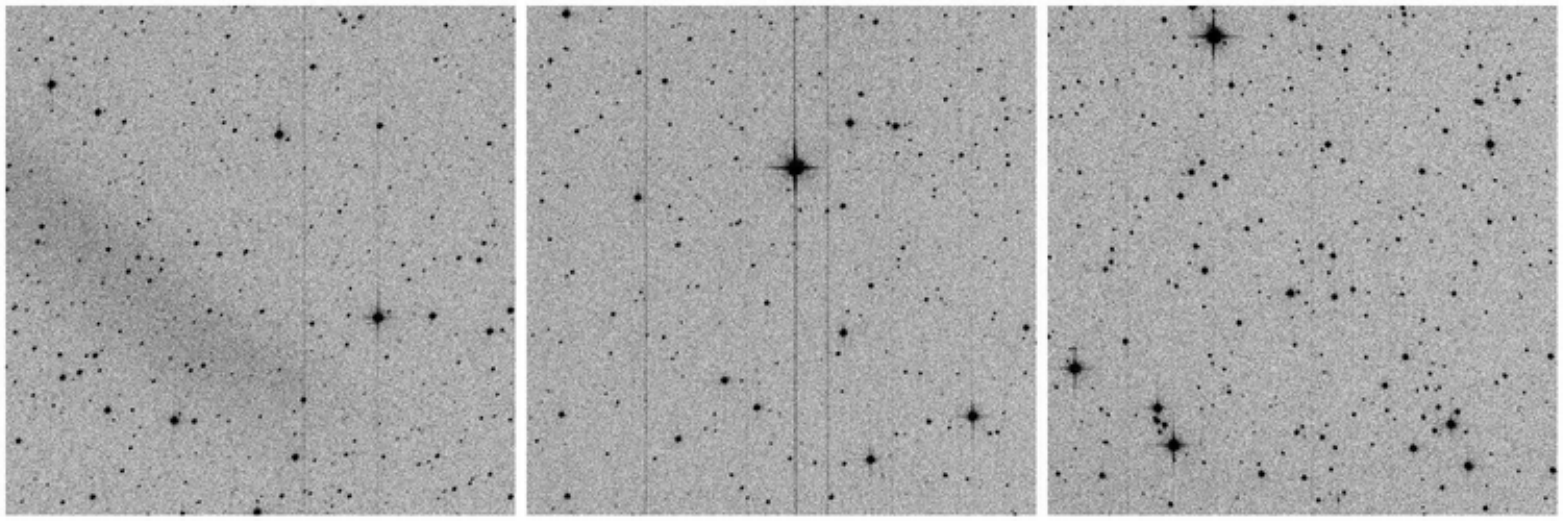}
        
    \caption{VT images obtained at angles of 10$^\circ$, 20$^\circ$ and 30$^\circ$ from the full Moon (left to right).}
    \label{fig:moon_10_20_30deg}
\end{figure}

Stray light suppression is critical for the VT's detection sensitivity. Commissioning tests confirmed that the VT meets its stray light requirements, limiting contributions to one-third of the sky background at 30 degrees from the full Moon (Figures \ref{fig:straylight} and \ref{fig:moon_10_20_30deg}). Notably, even at 20 degrees, the VT maintains its specified detection limits, demonstrating exceptional performance under bright lunar conditions.

Unlike ground-based telescopes, where atmospheric scattering of moonlight significantly degrades observations—even with well-controlled telescope stray light—the VT avoids this issue entirely, showcasing the inherent advantage of space-based observations.

\begin{figure}
    \centering
    \includegraphics[width=0.8\linewidth]{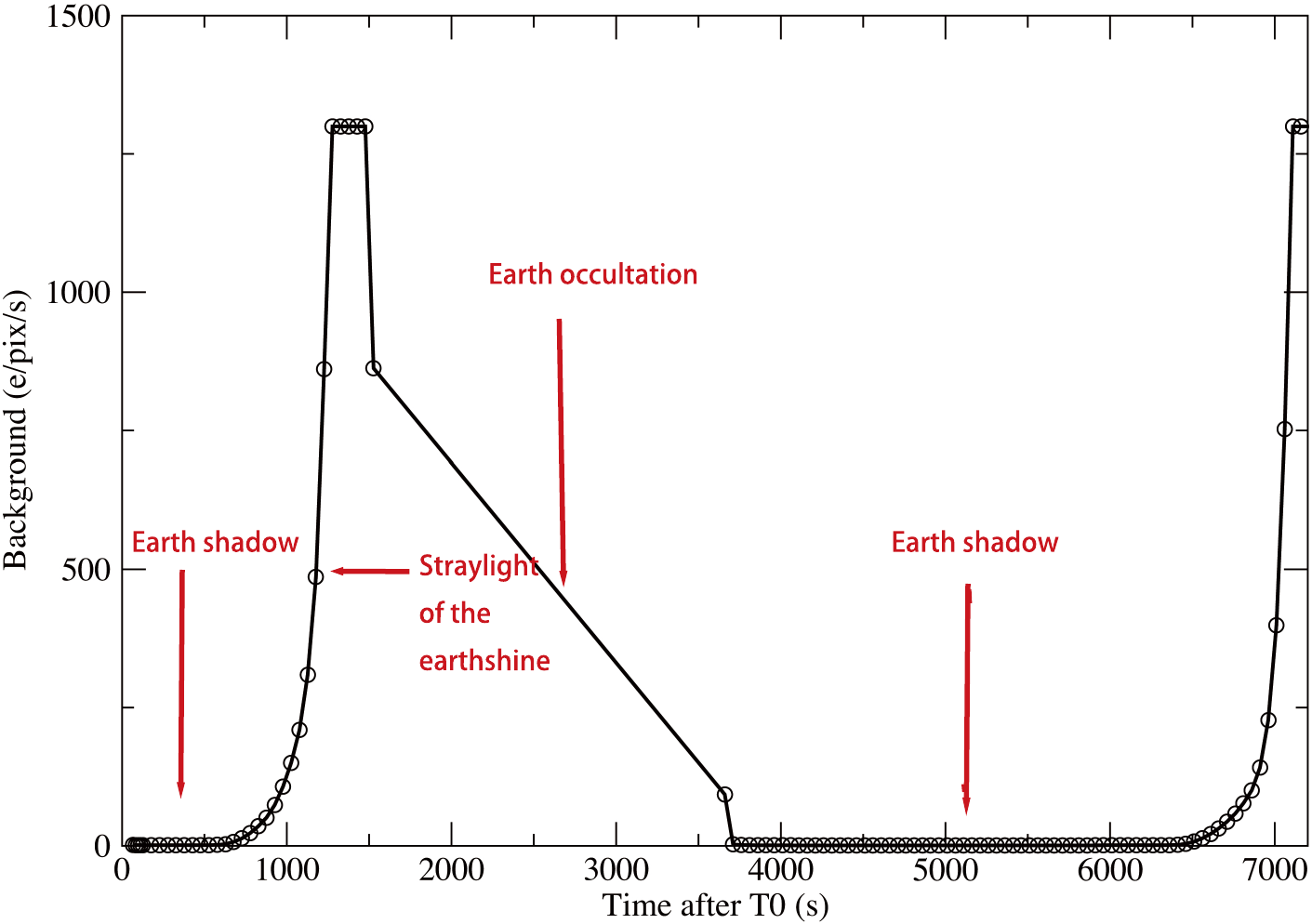}
    \caption{Evolution of the VT background over two orbital cycles, demonstrating the influence of Earthshine-induced stray light.}
    \label{fig:earth_shine}
\end{figure}

However, Earthshine dominates the VT’s stray-light background, as shown in Figure \ref{fig:earth_shine}  for two post-GRB 250806A orbits. The background flux remains at a very low level in Earth’s shadow but increases quickly upon exit, reducing the effective exposure time to 35  minutes per 96-minute orbit after accounting for occultation.

\subsection{Throughput}

Throughput monitoring was conducted using spectrophotometric standard stars (see \citealt{Yao+etal+2026} for details). Initial measurements during the commissioning phase indicated a 30\% flux deficit relative to ground-based calibrations, severely constraining detection limits. Subsequent testing ruled out shutter timing anomalies, indicating that the contamination—likely on the detector surfaces or protective windows—was the root cause. A de-contamination campaign, heating the protective windows from $-25^\circ$C to $10^\circ$C, mitigated the issue and recovered 20\% of the flux. The residual 10\% loss may stem from mirror reflectivity degradation over the two-year interval or persistent contamination on the surfaces of CCDs. The zero-point magnitude is therefore corrected for the observed flux loss, alongside ongoing monitoring of throughput and contamination effects.

\subsection{Platform pointing stability}
\label{sect:stability}

Platform pointing stability is critical for early GRB observations after slewing maneuvers, as image jitter can significantly degrade the PSF and reduce detection sensitivity. To ensure stability, \textit{SVOM} employs a Fine Guidance Sensor (FGS, \citet{Li+etal+2026a}) co-mounted with the VT's blue-channel CCD in the same focal plane, achieving sub-arcsec pointing precision. The FGS operates in a closed control loop with the Attitude and Orbit Control System (AOCS) to maintain precise pointing. 

The platform's pointing stability is evaluated by analyzing imaging shifts in VT under two operational scenarios: (1) short exposures during ToO observations, assessing high-frequency stability (e.g., jitter); and (2) long exposures in GRB follow-up observations after large-angle slews, verifying post-maneuver settling performance.

\subsubsection{Short exposure ToO scenario}
\begin{figure}
    \centering
    \includegraphics[width=0.8\linewidth]{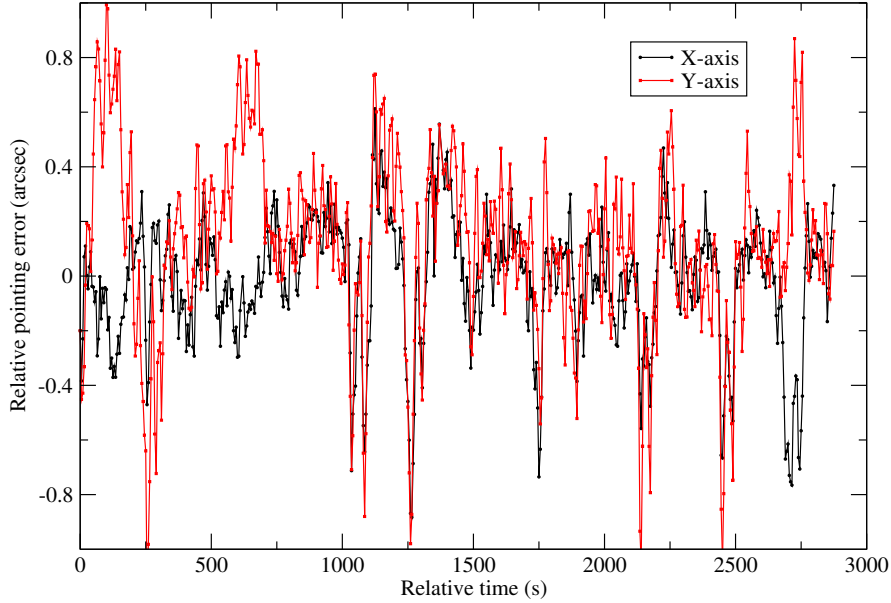}
    \caption{The figure presents the platform's stability during a ToO observation with a 5-second exposure, showing the relative pointing errors in both X and Y axes. }
    \label{fig:short_exposure_tracking}
\end{figure}

This study analyzes images from a ToO observation using 5-second exposures. To enable continuous imaging without dead time, the detector operated in windowed  readout mode (500 $\times$ 500), ensuring the readout time remained shorter than the exposure duration.  This study focuses on imaging shifts, as rotational effects are negligible (relative angular differences $<$0.005°, resulting in $<$0.1 arcsec shifts at the edges of VT images). Figure \ref{fig:short_exposure_tracking} demonstrates the platform's high-frequency stability during a full orbit of ToO observations, showing RMS residuals of 0.22" (X-axis) and 0.27" (Y-axis), which confirms excellent short-term pointing performance.

\subsubsection{Long exposure GRB scenario}

\begin{figure}

    \centering
    \includegraphics[width=0.8\linewidth]{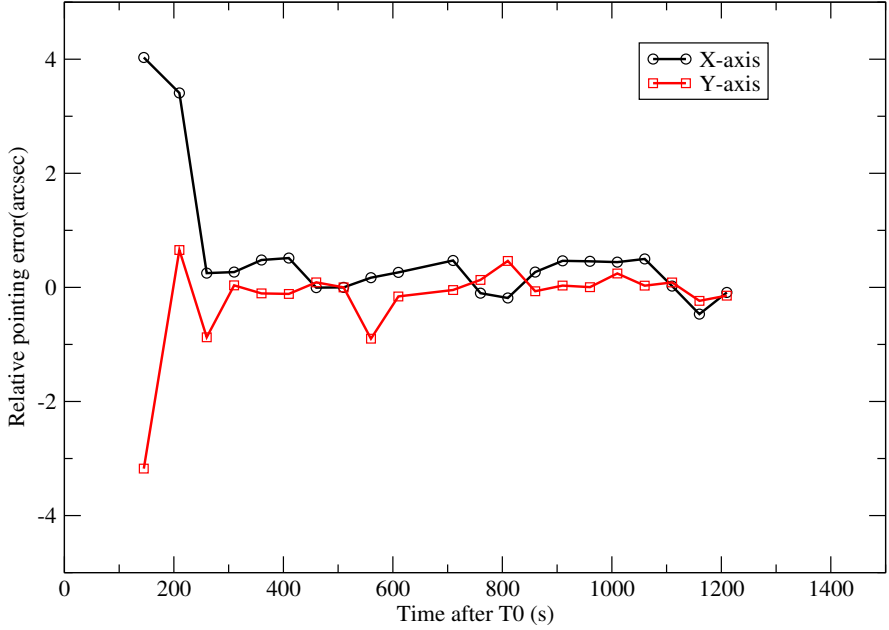}
    \caption{The figure presents the tracking performance after slewing to GRB 250806A, demonstrating the platform's stabilization dynamics and pointing accuracy achieved through the FGS in the AOS control loop during gamma-ray burst follow-up observations.}
    \label{fig:long_exposure_tracking}
\end{figure}

The long-exposure tracking performance was evaluated using follow-up observations of GRB 250806A, with 50-second exposures per frame. The analysis focused on images acquired from the end of a large-angle slew maneuver to the onset of Earth occultation in the first post-burst orbit.  Figure \ref{fig:long_exposure_tracking} shows the post-slew centroid displacements along both X- and Y-axes. The initial measurement, taken immediately after slew completion and successful target acquisition in the VT's FOV, captures the platform's transient unstable phase. Through FGS-assisted stabilization, the system attains preliminary pointing stability within 60 seconds post-maneuver, achieving full stabilization by 200 seconds after T0 with RMS residuals of 0.24" (X-axis) and 0.12" (Y-axis).

\subsection{Detection limits}

\begin{figure}

    \centering
    \includegraphics[width=0.8\linewidth]{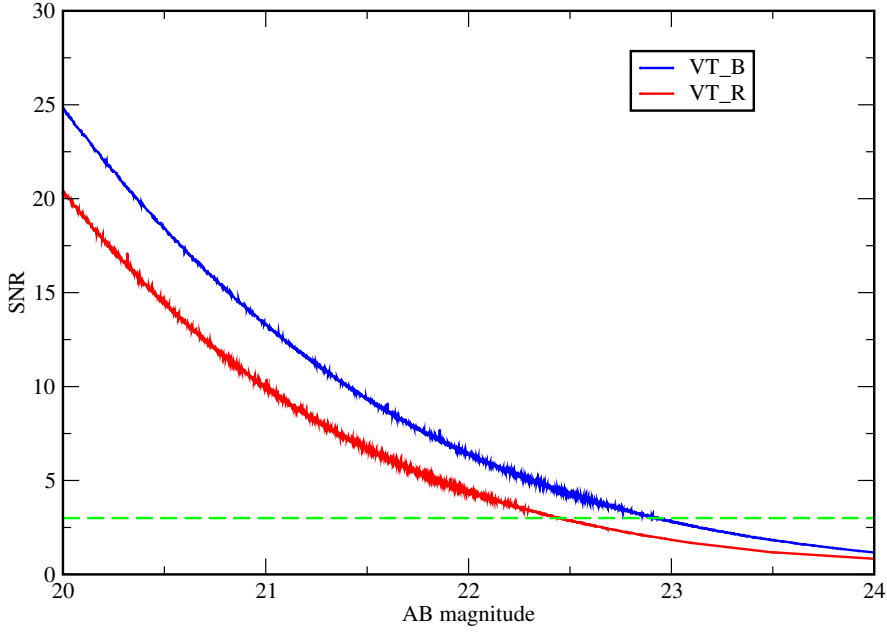}
    \caption{SNR versus AB magnitude for GRB 250314A. The $3\sigma$ detection limits, indicated by horizontal dashed lines, correspond to AB magnitudes of 22.8 (blue band) and 22.4 (red band).\vspace{8mm}}
    \label{fig:snr_ABmag}
\end{figure}

In-orbit performance confirms that the VT achieves its design requirement of a $3\sigma$ limiting magnitude of $M_V = 22.5$ in a 300-s exposure \citep{Zhang+etal+2026}. To validate these results, we analyze the signal-to-noise ratio (SNR) as a function of AB magnitude for the GRB 250314A field using a 6-frame stack (effective integration time: 300 s). The $3\sigma$ detection limits reach AB = 22.8 (blue band) and AB = 22.4 (red band) (Figure \ref{fig:snr_ABmag}). Further co-adding single-orbit observations enhances the sensitivity to AB $\sim 23.5$, while stacking three orbits pushes the limit to$\sim 24.0$.

\subsection{Co-alignment with Other Instruments}

Co-alignment calibrations between the VT and star trackers, as well as between the VT and MXT, were performed during the commissioning phase. Although the \textit{SVOM} instruments were coarsely aligned before launch, precise in-orbit calibration is essential to enable their synergy in GRB localization, given the lack of precise calibration sources on the ground. Throughout the mission, bias metrics are continuously monitored and alignment parameters updated if significant deviations between instruments are detected. 

\subsubsection{Alignment calibration between VT and star trackers}

The \textit{SVOM} satellite's optical axis is defined as the center of the VT FOV in the red band. To minimize thermo-elastic effects, the conversion from the platform's quaternions to the VT attitude matrix was performed using multiple pointing measurements, with the final matrix derived as the average of these measurements. 

\begin{figure}
    \centering
    \includegraphics[width=0.8\linewidth]{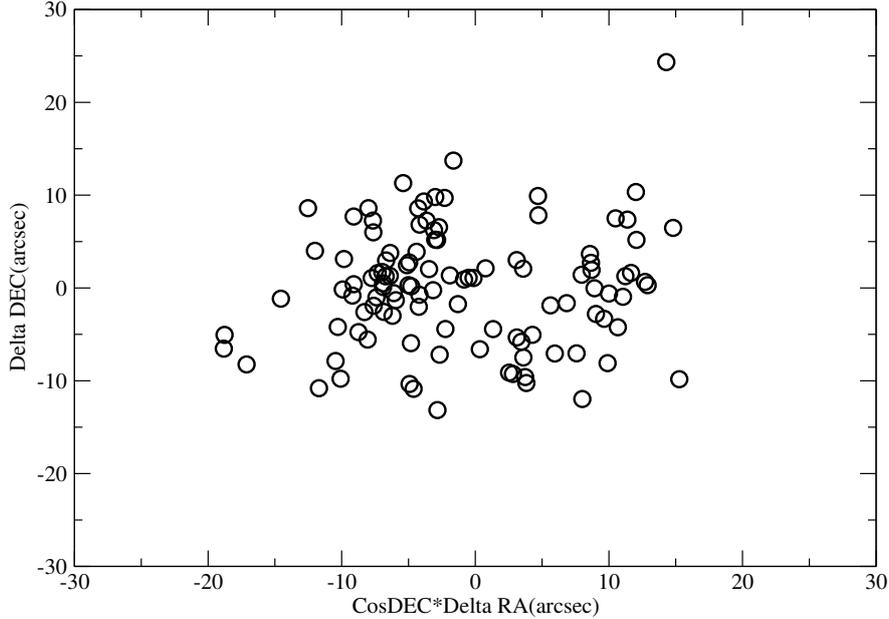}
    \caption{Residuals of the bias calibration between VT and star trackers.}
    \label{fig:VT_startracker}
\end{figure}

Figure \ref{fig:VT_startracker} shows the residuals from the calibration of 110 GRB fields observed by VT during the first year. The mean offsets and RMS residuals are -0.02 and 5.8 arcsec in Right Ascension (RA), and -1.0 and 7.7 arcsec in Declination (Dec), respectively. These results indicate good post-calibration co-alignment between the VT and star trackers.

\subsubsection{Alignment calibration between VT and MXT}

Accurate calibration of the alignment between \textit{SVOM}’s narrow-field telescopes, the VT and MXT, is essential for VT data processing and improved MXT localization, as the two instruments work synergistically in GRB observations. While the VT’s FOV can cover ECLAIRS’ localization error box, onboard processing relies on small windows centered on MXT’s pointing position to reduce VHF data downlink, necessitating precise alignment of the MXT optical axis within the VT focal plane. Conversely, the VT may offer more precise attitude measurements than star trackers, aiding the MXT in minimizing systematic errors for J2000 localization.

The calibrated MXT-VT bias matrix is detailed in \citet{gotz+etal+2026} and \citet{robinet+etal+2026}, with systematic errors in the matrix fitting constrained to better than 1 arcmin. Given the typical VT window size of 6 arcmin × 6 arcmin, this calibration accuracy is sufficient for the intended requirements. 

When transforming localization from the MXT frame to the VT frame using the calibrated matrix, only the angular coordinates ($\theta$, $\phi$) are obtained. To map these coordinates to the corresponding window position on the detector plane, an onboard conversion to pixel coordinates is required. This involves projecting the spherical coordinates onto a plane and then transforming them into pixel coordinates.

Ground calibration derived the fitting parameters for this conversion, which were uplinked to the satellite via telecommand. The resulting accuracy is at the arcsec level—negligible compared to the errors introduced by the MXT-VT bias matrix.

\section{Observational results during the first year}
\label{sect:resu}

The VT has been routinely performing autonomous follow-up observations and telecommanded ToO monitoring of nearly all GRBs triggered by ECLAIRS, as well as GRBs and other transients reported by \textit{Swift} and \textit{EP}. We present two cases to illustrate the VT’s capabilities during the commissioning and operational phases: the first ECLAIRS trigger to which the VT responded onboard, GRB 241018A, and the highest-redshift \textit{SVOM} GRB to date, GRB 250314A. Finally, we present a preliminary statistical analysis of GRB detections by the VT during the first year.

\subsection{GRB 241018A: The first VT GRB detected via onboard data processing}
\subsubsection{General information on GRB 241018A}

GRB 241018A was detected by ECLAIRS \citep{2024GCN.37812....1S} on 2024 October 18 11:54:34 UTC during the \textit{SVOM} commissioning phase. This bright, long-duration GRB was initially localized to within a 2 arcmin error radius. Notably, it triggered the first successful platform slew response to a GRB. The burst was subsequently detected by the MXT  following the slew, yielding a refined localization accuracy of 42 arcsec (with an estimated 1.5 arcmin systematic offset between the MXT  and VT coordinates).

\subsubsection{VT Follow-up observations}

The VT began observations at 11:56:36 UTC (T0+122 s) in chance mode with a sequence of 15 s short exposures. Following platform stabilization (meeting the operational threshold), it transitioned to standard 100 s exposures at 11:58:48 UTC. Observations during the first two orbits were performed in full-frame mode, after which the VT switched to 800 × 800-pixel window mode, centered on the MXT-derived localization in the VT reference frame.

\subsubsection{VT VHF data processing}

Onboard data processing was performed on four sequences of VT continuous observations. The first sequence began at 12:00:28.0 UTC, consisting of three consecutive images. The second sequence followed immediately, while the third and fourth sequences covered the second orbit, each comprising six images. Dark and flat-field corrections were applied to each image onboard, with every three consecutive images per channel subsequently stacked into a master image. A subframe was extracted from the master image, centered on the MXT localization, with dimensions scaled according to the MXT localization uncertainty.  Finally, both an FDC (the source list) and a 1-bit image were generated  and downlinked to the ground via the VHF channel for rapid data analysis.

Since the GRB occurred during the commissioning phase, the platform’s performance was suboptimal, and platform instability resulted in image jitter for some exposures. Consequently, the VT successfully generated source lists only for the sequences 1 and 3  and only two 1-bit images in the blue channel.

\begin{figure}
    \centering
    \includegraphics[width=0.8\linewidth]{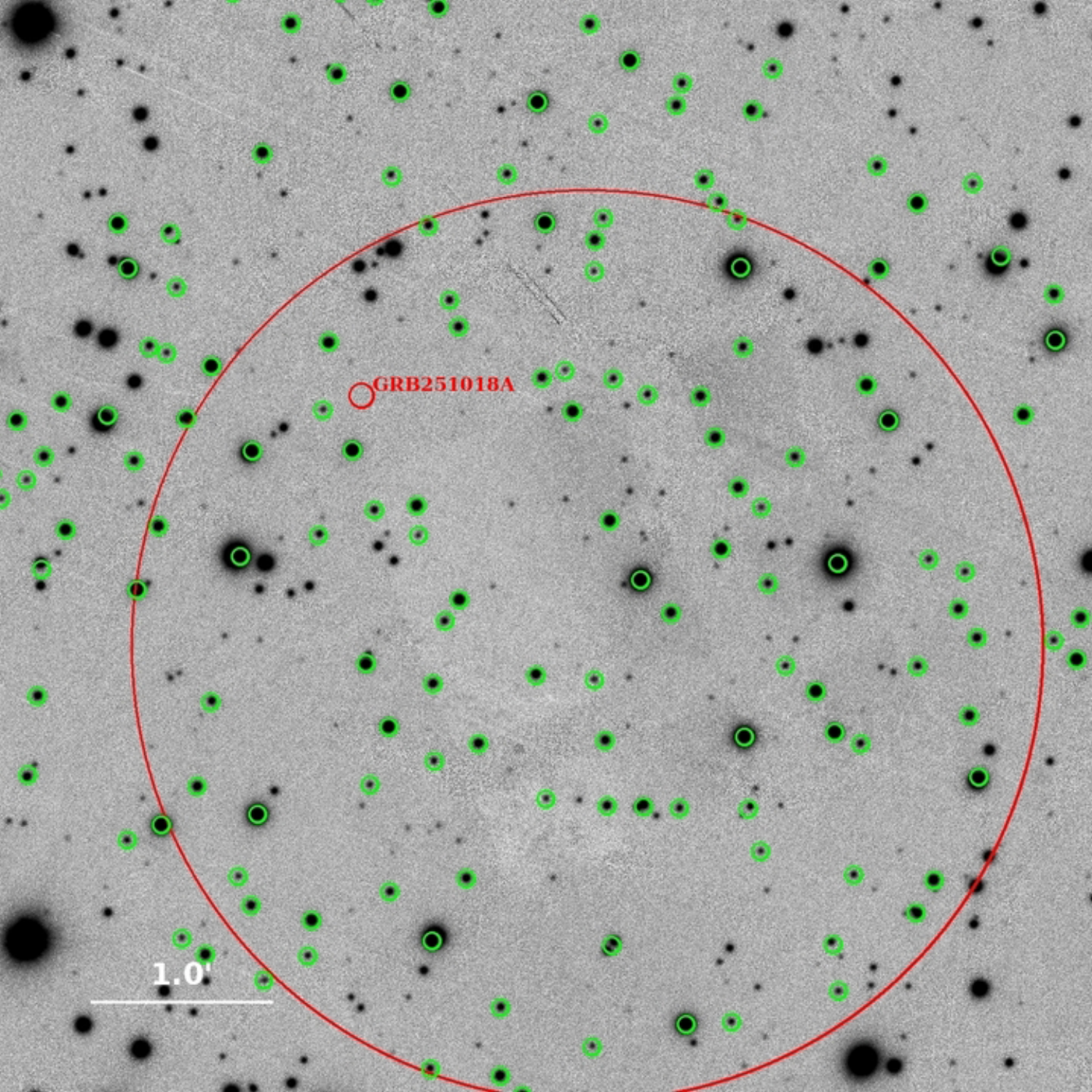}
    \caption{The uncatalogued optical counterpart candidate of GRB 241018A (small red circle) is shown superimposed on a Pan-STARRS g-band image. Reference sources (green circles) and the MXT localization region (large red circle; 2 arcmin radius) are also marked for comparison.}
    \label{fig:sourcelist_candidate}
\end{figure}

Following photometric and astrometric calibration using the ground pipeline (VVPP), cross-matching with existing catalogs identified an uncatalogued source in the FDC across both bands (see Fig. \ref{fig:sourcelist_candidate}). The candidate exhibited a $\sim$1 magnitude fade between sequences 1 and 3, strongly supporting its identification as the GRB's optical counterpart. These findings were promptly reported in a GCN Circular \citep{2024GCN.37819....1S}.

\subsubsection{VT X-band data processing}

\begin{figure}
    \centering
    \includegraphics[width=0.8\linewidth]{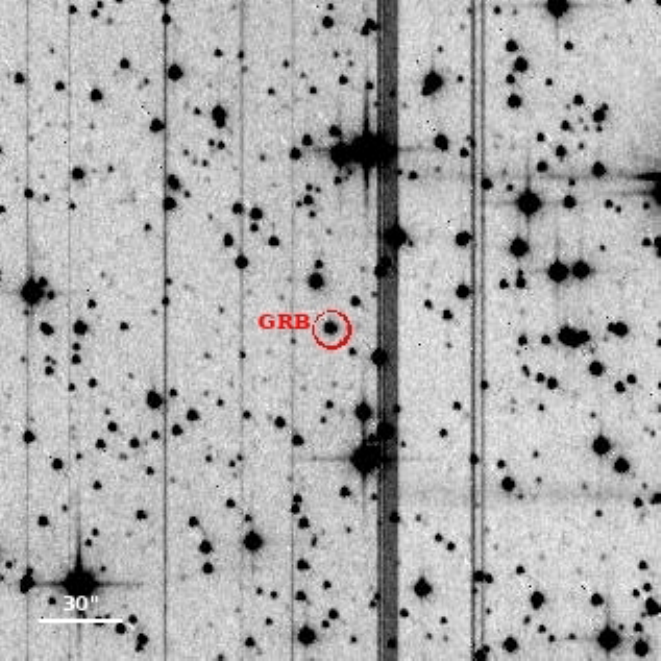}
    \caption{A combined VT image of GRB241018A (red circle) in the red band.}
    \label{fig:image_GRB241018A}
\end{figure}

\begin{figure}
    \centering
    \includegraphics[width=0.8\linewidth]{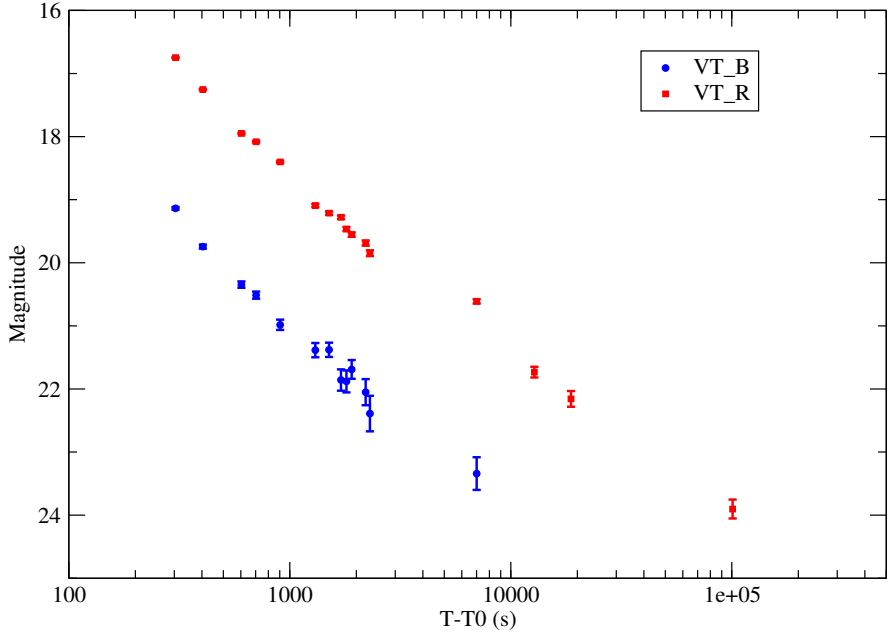}
    \caption{VT light curves of GRB 241018A for both blue and red bands.}
    \label{fig:lightcurves_GRB241018A}
\end{figure}

The candidate was reconfirmed using the complete raw data downlinked via the X-band and processed with the VXPP pipeline (see Fig. \ref{fig:image_GRB241018A} and \citealt{2024GCN.37826....1S}). Comprehensive light curves were then constructed for both VT bands (Fig. \ref{fig:lightcurves_GRB241018A}). 

This GRB provided a rare opportunity to evaluate the VT's stray light suppression under extreme conditions, occurring during a full Moon. Despite its proximity (34.1°) to the Moon, observations within Earth's shadow showed remarkably low background levels, demonstrating the instrument's robust stray light control. However, upon exiting the shadow, a sudden two-order-of-magnitude background surge occurred within the first 400 s—a clear signature of Earthshine-induced stray light (see Section \ref{sect:staylight}).

The VT's identification strategy for GRB optical counterparts was first validated by this event. Its performance has been further enhanced through subsequent pipeline optimizations, achieving rapid GRB localization with a precision of $<0.5$ arcsec, demonstrating efficient synergy with ECLAIRS and MXT.

Although no spectroscopic redshift is available for this GRB, we have nevertheless estimated rough redshift limits using the observed VT color index between the blue and red bands which is approximately 2 magnitudes. Due to its low Galactic latitude ($b = -3.6^\circ$), a Galactic dust extinction correction of $E(B-V) \approx 0.66$ was applied, yielding a dereddened color index of $\approx 1.5$. According to the statistical relationship between the VT color index and GRB redshift synthesized by  \citep{2020RAA....20..124W}, this event likely occurred at a low to intermediate redshift ($z<4$).

\subsection{GRB 250314A: \textit{SVOM}'s highest-redshift GRB to date}

\textit{SVOM} detected five GRBs with redshift $z>4$ in the first year, including GRB 241025A (z = 4.20) \citep{2024GCN.37866....1X}, GRB 251003A (z = 4.41) \citep{2025GCN.42086....1S},  \textit{EP}250215a/GRB 250215A (z = 4.61) \citep{2025GCN.39343....1S}, GRB 250725A (z = 5.26) \citep{2025GCN.41160....1T}, and GRB 250314A (z = 7.3) \citep{2025GCN.39732....1M}. Among these, GRB 250314A is the highest-redshift burst detected by \textit{SVOM} to date. Complementing the brief discussion of its VT follow-up observations in \citet{2025arXiv250718783C}, which was limited by scope constraints, we present here a detailed analysis of the VHF and X-band VT data to elaborate on prior findings.

\subsubsection{General information on GRB 250314A and VT follow-up observations.}

GRB 250314A \citep{2025GCN.39719....1W} was detected by ECLAIRS on 2025 March 14 at 12:56:42 UTC with a SNR of 9.1, triggering a platform slew at T0+27 s; it was also recorded by the GRM. Like GRB 241018A, it also occurred during a full Moon and was located only 25 degrees away from the Moon. While GRB 241018A occurred during SVOM’s commissioning phase, this burst was detected two months into the operational phase. By then, both the spacecraft platform and the VT had stabilized, performing nominally and meeting all design specifications.

VT observations commenced at 12:59:03 UTC (141 s after T0) , initiating with a sequence of 15 s exposures in chance mode. After 45 s, once stability criteria were satisfied, the instrument transitioned to a 50 s long-exposure mode, which persisted until the end of the second orbit. Observations resumed in the third orbit with 100 s exposures, but these were interrupted at 16:17:49 UTC by a secondary slew-triggering event (sb25031408). Subsequent ground analysis confirmed this trigger as false.

Due to the absence of an MXT localization, data were recorded in full-frame mode, though only every other frames were retained to optimize downlink bandwidth.

\subsubsection{VT VHF data processing}

Onboard processing was conducted on four exposure sequences during the first two orbits, with processed data downlinked via VHF. The first sequence began at 13:01:28 UTC (T0+284 s) and consisted of three stacked frames with a total integration time of 150 s. A second sequence followed at 13:03:58 UTC, featuring six frames spanning 300 s. Before Earth occultation, VT extracted and processed a 10 arcmin  × 10 arcmin window centered on the FOV, following a predefined strategy in the absence of prior MXT detections. The third and fourth sequences during the second orbit adopted the same window size and position due to the non-detection by the MXT.

\begin{figure}
    \centering
    \includegraphics[width=0.8\linewidth]{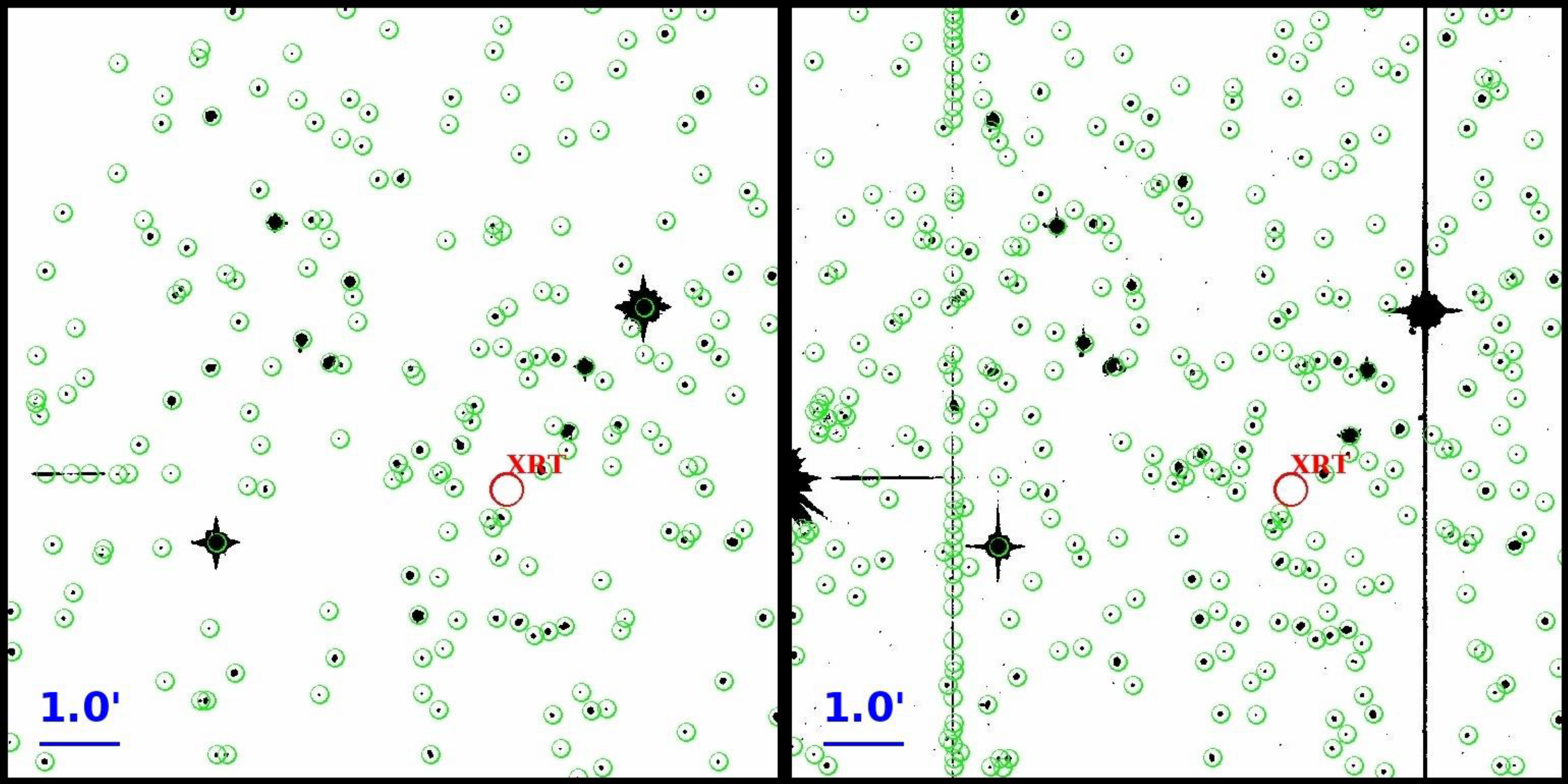}
    \caption{
    VT 1-bit images ($770 \times 770 $ pix., $10\arcmin \times 10\arcmin$ fov) of GRB 250314A in the blue (left) and red (right) bands, highlighting FDC sources (green circles) and the Swift/XRT localization (red circle, 10 arcsec radius).}
    \label{fig:GRB250314A_VHF_1bit}
\end{figure}

Figure \ref{fig:GRB250314A_VHF_1bit} displays the 1-bit images for both VT bands derived from onboard processing of the first sequence data, overlaid with the FDC source lists. No counterpart was detected in the VHF data. While a limiting magnitude of 20 was reported in our initial VHF results \citep{2025GCN.39722....1P}, subsequent refined analysis of the FDCs, following the detection of the X-ray afterglow by \textit{Swift}/XRT \citep{2025GCN.39734....1K}, yielded a deeper limit of 22 mag.

\subsubsection{VT X-band data processing}

\begin{figure}
    \centering
    \includegraphics[width=0.8\linewidth]{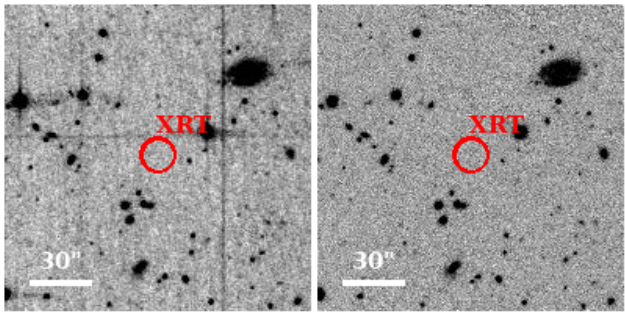}
    \caption{
    VT red-band image (left) and Legacy Survey r-band image (right) of the GRB 250314A field (2.5 arcmin × 2.5 arcmin). Red circles show the XRT localization (10 arcsec radius).}
    \label{fig:GRB250314A_xband_image}
\end{figure}

Following the acquisition of X-band data, we carried out a detailed analysis. The stacking of the first 12 stable VT images (50 s exposures each) produced a limiting magnitude of 23, yet no source was identified within the Swift/XRT error region. The stacked VT red-band image is displayed in Figure \ref{fig:GRB250314A_xband_image} alongside a DESI Legacy Survey \citep{desi_lagacy_survey} r-band image, both covering the same 2.5 arcmin × 2.5 arcmin field.

The early report of the VT’s non-detection—with an exceptionally deep limiting magnitude—prompted near-infrared follow-up observations with the NOT telescope \citep{2025GCN.39727....1M}. These observations led to the identification of a high-redshift GRB, whose redshift (z = 7.3) was later measured via spectroscopic observations with the VLT \citep{2025GCN.39732....1M}.

\subsection{First-Year GRB Detection Rate of the VT}

\subsubsection{Observations of \textit{SVOM} GRBs}

Ground-based identification and photometry of SVOM GRBs are conducted using either early-time onboard pipeline outputs (Section \ref{sect:vvpp}) or late-time X-band downlinked data (Section \ref{sect:vxpp}). The detection statistics for both sources are briefly presented below.

\textbf{Prompt Analysis of Onboard Pipeline Outputs}
As of 2025 September 3, SVOM/ECLAIRS had detected 32 gamma-ray bursts (GRBs) with prompt slew observations, of which 25 were successfully processed onboard. Some bursts were detected during the commissioning phase, when both the platform and VT were not yet operating at full performance. The first data were downlinked to the ground via VHF with a median latency of 18 minutes post-trigger. Optical counterparts were identified for 14 bursts, yielding a 56\% detection rate. This high early-time identification efficiency—enabled by the VT’s onboard processing—allows for rapid redshift measurements by ground-based telescopes during the bright afterglow phase. Complete statistical analyses are presented in \citet{wu+etal+2026}.

\textbf{Comprehensive Analysis of X-Band Data Products} By 2025 December 3, the VT had observed 43 ECLAIRS-triggered GRBs with auto-slew of the platform, achieving the observations with effective start times ranging from several minutes to 2 hours. X-band data analysis confirmed optical counterparts for 33 bursts, yielding a 76\% overall detection rate. For the subset of bursts with observation start times $\leq$10 minutes (21 bursts), the detection rate increased to 85\% (18 bursts)—more than twice the efficiency of Swift/UVOT ($\sim40\%$).  Detailed statistical analyses for VT detections of \textit{SVOM} bursts are presented in \citet{Li+etal+2026b}. 

This high-detection-rate sample enables systematic studies of dark GRBs. Additionally, continuous multi-wavelength monitoring—spanning minutes to hours post-trigger—revealed complex light-curve evolution, including early-time flares, chromatic variability (indicating spectral evolution), and late-time rebrightening, suggestive of an emerging supernova component. These findings are currently under detailed analysis and will be published in the coming months.

\subsubsection{ToO Follow-up Observations of SVOM Subthreshold GRBs and External Mission Triggers}

For SVOM GRBs below the detection threshold, the absence of an automated slew results in delayed VT observations, requiring ToO follow-up. As of 2025 December 3, the VT had detected 12 out of 19 targeted subthreshold bursts (63\%), showing a lower detection efficiency compared to automatically triggered slews. This reduced success rate likely stems from both the intrinsically lower fluence of these bursts and the increased observational latency.

Beyond its high detection efficiency for auto-slewed SVOM GRBs, the VT has proven equally capable in observing external mission triggers (e.g., \textit{Swift/EP}) via ToO requests. As of the same date, the VT had observed 49 external GRBs, detecting 38 events  with a 78\% detection rate.
Notably, since 2025 May, the BeiDou uplink has served as SVOM's primary transmission pathway for ToO commands, reducing median response latency from 13.8 hours to 8 hours—a 42\% improvement. This enhancement has demonstrably increased GRB detection efficiency. A detailed investigation of VT's detection efficiency and associated statistical properties appears in \citet{Li+etal+2026b}.
  
\section{Conclusions and perspectives}
\label{sect:summ}

The VT aboard the \textit{SVOM} satellite plays a key role in the optical identification of GRBs triggered by ECLAIRS, while also demonstrating strong synergy with other high-energy missions such as \textit{Swift} and \textit{EP} through its deep sensitivity and rapid follow-up capabilities. Performance tests confirm the VT effectively controls contamination and meets stringent design specifications, achieving sensitivities of 22.5 AB mag (300-s exposure) in early GRB phases and  $\sim 24$ AB mag (1.5-hour cumulative exposure) in later phases. Its fully operational onboard and ground-based pipelines, supported by rapid VHF downlinks, ensure timely optical counterpart identification and efficient data dissemination to the community.

The VT achieves an impressive detection efficiency of $\sim85\%$ for SVOM-triggered GRBs when observations begin within 10 minutes of the alert. For \textit{Swift/EP}-detected bursts, it maintains a comparable efficiency ($\sim78\%$) even with follow-up delays of several hours—nearly doubling the detection rate of \textit{Swift}/UVOT ($\sim\!40\%$). This exceptional performance underscores the VT's robust capability in GRB counterpart detection, solidifying its role as a critical instrument for the SVOM mission.

Beyond its high detection efficiency, the VT has also made significant contributions to the study of high-redshift GRBs. A prime example is its rapid response to GRB 250314A (z = 7.3), where the VT provided early, deep optical upper limits (up to $\sim 1\mu m$) around the Swift/XRT localization. These observations delivered critical constraints, strongly suggesting a high-redshift origin and triggering rapid follow-up with large ground-based telescopes equipped for NIR imaging. 

Looking ahead, the VT is well-positioned to detect and characterize a larger, scientifically robust sample of GRBs, with an emphasis on high-redshift events. Through deep, systematic monitoring, it will deliver essential insights into GRB central engines, circumburst environments, jet kinematics, and the origins of optically dark bursts.

\begin{acknowledgements}
The Space-based multi-band astronomical Variable Objects Monitor (\textit{SVOM}) is a joint Chinese-French mission led by the Chinese National Space Administration (CNSA), the French Space Agency (CNES), and the Chinese Academy of Sciences (CAS). We gratefully acknowledge the unwavering support of NSSC, IAMCAS, XIOPM, NAOC, IHep, CNES, CEA, and CNRS. This work is supported by the Strategic Priority Research Program of the Chinese Academy of Sciences (Grant No.XDB0550401), and by the National Natural Science Foundation of China (grant Nos. 12494571 and 12494570, 12494573, 12133003). The authors are thankful for support from the National Key R\&D Program of China (grant Nos. 2024YFA161170* and 2024YFA1611700). 
\end{acknowledgements}

\bibliography{ms2026_0014}{}

\label{lastpage}
\clearpage

\end{document}